\documentstyle[12pt]{article}

\def\ov{\overline}
\def\s2{\frac{1}{\sqrt2}}
\def\wh{\widehat}
\def\oh{\frac{1}{2}}
\def\sh{\frac{s}{2}}
\def\IZ{Z\kern-.4em  Z}
\topmargin
-1.0cm
\textwidth
15cm
\textheight
22.5cm
\oddsidemargin
1cm
\evensidemargin
1cm
\begin{document}
\newcommand{\newc}{\newcommand}
\newc{\sm}{Standard Model}
\date{}
\title{ Building GUTs from Strings}

\author{G. Aldazabal \thanks{Permanent Institutions: CNEA, Centro At\'omico
Bariloche, 8400 S.C. de Bariloche, and CONICET, Argentina} $^1$,
A. Font$^2$, L.E. Ib\'a\~nez$^1$ and A. M. Uranga$^1$ \\ \\
$^1$Departamento de F\'{\i}sica Te\'orica, \\
Universidad Aut\'onoma de Madrid, \\
Cantoblanco, 28049 Madrid, Spain.
\\   \\
$^2$Departamento de F\'{\i}sica, Facultad de Ciencias,\\
Universidad Central de Venezuela,\\
A.P. 20513, Caracas 1020-A, Venezuela.\\
}
\maketitle
\vspace{-4in}\hspace{4in} FTUAM-95-27

\hspace{4in}
hep-th
\vspace{4in}
\begin{abstract}

We study in  detail the structure of Grand Unified
Theories derived as the low-energy limit of orbifold
four-dimensional strings.
To this aim, new techniques for building
level-two symmetric orbifold theories are presented.
New classes of GUTs in
the context of symmetric orbifolds are then constructed.
The method of permutation modding is further explored and
$SO(10)$ GUTs with both $45$
or $54$-plets are obtained. $SU(5)$ models are also found through this
method. It is shown that, in the
context of symmetric orbifold $SO(10)$ GUTs, only
a single GUT-Higgs, either a $54$ or a $45$, can be present and
it always resides in an order-two untwisted sector.
Very restrictive results also hold in the case of $SU(5)$.
General properties and selection rules for string GUTs
are described. Some of these selection rules
forbid the presence of some particular GUT-Higgs couplings which are
sometimes used in SUSY-GUT model building. Some semi-realistic
string GUT examples are presented and their properties briefly discussed.

\end{abstract}
\maketitle

\newpage
\section{Introduction}

In a recent paper \cite{afiu}, we discussed the construction of
four-dimensional string models whose massless sector constitute grand unified
$SO(10)$ and $SU(5)$ theories. In spite of the popularity of both
four-dimensional strings and GUT ideas, it is surprising how
little effort has been devoted to making compatible both approaches
to unification by constructing GUTs from strings. The reason for this is that
standard GUTs require the presence of GUT-Higgs chiral fields, e.g. adjoints in
$SU(5)$, adjoints or $54$s in $SO(10)$, in the massless spectrum. In order to
have that type of massless chiral fields in a chiral $N=1$ theory,
the affine Lie algebra associated to the GUT group has to be realized
at level two or higher. Straightforward compactifications of the
supersymmetric heterotic string always have level-one algebras inherited from
the $E_8\times E_8$ or $SO(32)$ ten-dimensional heterotic string.
To obtain 4-d strings with higher level it is necessary to go beyond simple
compactifications of the heteorotic string. At the beginning it was thought
that such higher level models would be very complicated to construct.
This explains why,
in the early days of string model-building, there were no attempts in this
direction. In fact, only a few papers have dealt with the explicit construction
of 4-d strings with affine Lie algebras at higher levels
\cite{lew,fiq,lyk,cleav}.

Actually, it turns out that it is not particularly complicated to construct
higher level models and hence 4-d strings whose massless
sector constitute $SU(5)$ and $SO(10)$ GUTs. In ref.\,\cite{afiu}, we
referred to this type of models as string GUTs and investigated their
construction in the framework of symmetric Abelian orbifolds.
Abelian toroidal orbifolds with $N=1$ unbroken supersymmetry
are probably the simplest non-trivial 4-d strings
and hence provide a natural context for a general study of
string GUTs. Furthermore, consistency with world-sheet supersymmetry,
which is often a problem \cite{lyk,cleav}
in the alternative fermionic construction, is
built-in in the orbifold formalism.

The general prescription to build higher level models is the following.
The starting point is a $(0,2)$ orbifold compactification of the 10-d
heterotic string. It turns out convenient to use the $Spin(32)/Z_2$
instead of the $E_8\times E_8$ lattice. Models in which the gauge
group has the structure $G_{GUT}\times G$, where $G$ in turn contains
as a subgroup a copy of $G_{GUT}$, i.e. $G_{GUT}\subset  G$, are searched for.
At this point we have an usual level-one (0,2) orbifold model with a
particular gauge structure. The next step is to perform a modding or
projection such that only gauge bosons corresponding to the {\it diagonal}
$G_{GUT}^D$ subgroup of $G_{GUT}\times G_{GUT}$ survive in the massless
spectrum. We are thus left with a gauge symmetry $G_{GUT}^D\times G'$
in which $G_{GUT}^D$ is realized at level two.

In the case of orbifolds, the final reduction
$G_{GUT}\times G_{GUT}\rightarrow G_{GUT}^D$ can be achieved following
three different methods explained in refs.\,\cite{afiu, fiq}. In method I,
the underlying level-one model with $G_{GUT}\times G$ group is obtained
by embedding the twist action of the orbifold into the gauge degrees of
freedom by means of an automorphism of the gauge lattice instead of a shift.
Then, a ``continuous Wilson-line'' \cite{inq2} background is added in
such a way that the symmetry is broken continuously to the diagonal subgroup
$G_{GUT}^D$. In method II, the final step is a
modding of the original model by a $Z_2$ twist under which the two factors of
$G_{GUT}$ are explicitly permuted. Method III
is field-theoretical, the original symmetry is broken down to the diagonal
subgroup by means of an ordinary Higgs mechanism.

Although the above three methods in principle look different, there are
many level-two models that can be built equivalently using more than
one of them. In fact, the scheme is quite general and may be implemented
in other classes of 4-d string constructions. For example,
it may be used within the class of $(0,2)$ models \cite{mond,fiqs}
obtained by adding gauge backgrounds and/or discrete torsion to Gepner and
Kazama-Suzuki models. The final step leading to a level-two
group can be achieved by embedding an internal order-two symmetry
into a permutation of the two $G_{GUT}$ factors.

In the present paper we report on a number of new results obtained in
the construction of string GUTs from symmetric orbifolds.
These new results are discussed in the different chapters
as follows.

In chapter two we describe in detail the conditions for
modular invariance in the method in which a permutation is used to
obtain the level-two GUT. This is done by a careful study of the
partition function. We undertake this analysis of the permutation method
for two reasons. Modding by permutation of gauge coordinates has not been
studied in any detail in the literature and there are several non-trivial
technical issues to be fixed before the actual construction of GUT
models. Moreover, unlike in the other two methods, the resulting
examples are not necessarily continuously connected to an original
level-one theory. In chapter three we then apply our analysis to enlarge our
class of models and construct examples that we were unable to discover before.
In particular, considering the simultaneous action of a shift and a
permutation in the gauge lattice, allows us to derive the first symmetric
orbifold $SO(10)$ GUTs with an adjoint in the massless spectrum. We also
obtain $SU(5)$ models using the permutation plus shift method. In contrast,
in \cite{afiu}, the $SO(10)$ GUTs contained only $54$s and $SU(5)$ examples
were only constructed using the method of flat directions.

In chapter four we further explore the general structure of $SO(10)$ and
$SU(5)$ symmetric orbifold string GUTs. Expanding our analysis of
ref.\,\cite{afiu}
we show that this class of $SO(10)$ string GUTs can only have
either one $54$ or one adjoint $45$ and that this GUT-Higgs can only surface
in the untwisted sector. Furthermore we show that only orbifolds with
one complex plane rotated by just order-two twists can lead to $SO(10)$
string GUTs. Abelian $Z_N$ and $Z_N \times Z_M$ orbifolds with this property
are the $Z_4$, $Z_6$, $Z_8$, $Z_{12}$, $Z_2\times Z_2$, $Z_2\times Z_4$
and $Z_2\times Z_6$ orbifolds. Hence, many symmetric orbifold models
are ruled out for the purpose of $SO(10)$ GUT-building.
In the case of $SU(5)$ the situation is slightly less tight. The adjoint
$24$ can belong either in untwisted sectors or in twisted sectors
corresponding to certain particular order four or six twists. Many orbifolds
such as $Z_3$, $Z_3\times Z_3$, $Z_7$, $Z_2\times Z_6'$ and $Z_8'$, are
thus unsuitable for string GUTs. When GUT Higgses do occur, their structure
is very restricted and implies a set of selection rules on couplings that
are also presented in chapter four. Other selection rules that apply to
general string GUTs, and not exclusively to symmetric orbifolds, are also
given.

In chapter five we briefly discuss four-generation models as well as
a model with {\it almost } three generations. As we explained above, only
orbifolds of even order may lead to string GUTs.
Due to this even order there is the tendency to find an even
number of generations and hence four generations is the simplest option.
Although a three-generation $SU(5)$ model is presented, it also contains
an extra exotic ($15+9\cdot\ov{5}$) chiral family.
We believe that obtaining three net generations within
the context of symmetric orbifold GUTs is going to be a difficult

task. However, we do not think that this is a general property of
string GUTs. It should be possible to obtain three-generation examples
using similar techniques but applied to
other constructions such as
those based on Gepner or Kazama-Suzuki models.

In chapter six we summarize our conclusions and discuss the outlook
of string GUTs.

\section{The Permutation Method  Revisited }

In this chapter we upgrade and complete the formalism
discussed in \cite{afiu} for the construction of level-two
models from a permutation modding of the gauge coordinates.
In ref.\,\cite{afiu}  we discussed three methods for obtaining level-two
symmetric orbifold string-GUTs. Two of them, the method of
continuous Wilson lines and the method of field theoretical flat
directions, are continuously connected to level-one models.
This means that at particular points of the scalar fields moduli
space the level-two gauge symmetry is enhanced to a (bigger)
level-one symmetry. In this sense the level-two model is more
generic than its level-one parent.

In the case of $SO(10)$ string GUTs this continuous connection implies,
as argued in \cite{afiu}, that only $54$s GUT Higgs fields can be
obtained. In order to be able to find adjoint $45$s we have to use
a method that yields models not necessarily continuously
connected to an underlying level-one model. Since the permutation
modding involves a discrete order-two twist, it offers this kind of
possibility, and this is one of the motivations for the detailed analysis
presented in this section. It turns out that to obtain
$45$ representations, we have to simultaneously act with a discrete
shift on the gauge degrees of freedom, as will be shown below.
The permutation modding method will also allow us to construct
directly $SU(5)$ string GUTs which we were only able to derive in
\cite{afiu} through the flat direction method.

The starting point is the ten-dimensional heterotic string with gauge group
$E_8 \times E_8$ or $SO(32)$ corresponding to an affine Lie algebra at level
$k=1$. We use the bosonic formulation in which the gauge group originates
in 16 left-moving coordinates $F_J$ compactified on a torus with $E_8 \times
E_8$ or $Spin(32)/Z_2$ lattice denoted $\Lambda_{16}$. An usual orbifold
compactification \cite{orb2,imnq}
leads to to a 4-d model with gauge group $G_1 \times
G_2 \times \cdots$ in which each non-Abelian factor realizes an affine algebra
still at level $k=1$.

To obtain higher level algebras, a further process of twisting must be
performed.
Indeed, the main ingredient in the permutation method is an order-two operation
$\Pi$ that exchanges the $F_J$ coordinates associated to two identical group
factors $G\times G$, thus producing the diagonal subgroup $G_D$ at level $k=2$.
In the orbifold construction, the permutation $\Pi$ must be accompanied by an
action on the internal degrees of freedom. In the following we wish to explain
how $\Pi$ is implemented as a quantized Wilson line.

Our strategy will be to construct a modular invariant partition function for
the
orbifold with the $\Pi$ action. To do this we carefully deduce the contribution
 of the
simply-twisted sector and then generate the remaining sectores by modular
transformations and by requiring an operator interpretation. We will develop
explicitly the case of the $Z_4$ orbifold on an $SU(4) \times SU(4)$ internal
lattice.  However, many of the statements will apply to other orbifolds. In
particular, at the end we will consider the $Z_2 \times Z_2$ orbifold.
These two are the simplest even-order Abelian orbifolds.

\subsection{Notation and Embedding}

We introduce our notation by recalling some basic facts about orbifolds. We
denote the internal twist generator by $\theta$ and the internal lattice by
$\Gamma$, its basis vectors are $e_i$.  The eigenvalues of $\theta$ are of
the form $e^{\pm 2\pi i v_a}$,  $a=1,2,3$, with $0 \leq |v_a| < 1$. The $v_a$
can be chosen so that $v_1+v_2+v_3=0$ and define a twist vector
$(v_1,v_2,v_3)$. For convenience we also set $v_0=0$.
The orbifold action on the
internal coordinates is represented by an space-group element
$(\theta, n_ie_i)$, meaning boundary conditions
\begin{equation}
x(\sigma+1 ,t) = \theta x(\sigma,t) + n_ie_i
\label{xbc}
\end{equation}
In turn this implies that in the simply-twisted sector, the string Hilbert
space
splits into sub-sectors in which the string center of mass sits at fixed points
$x_f$ satisfying $(1-\theta)x_f = n_ie_i \, \hbox{mod} \, \Gamma$.  In each
$SU(4)$ sub-lattice there are four distinct choices for $n_ie_i$,  namely
$n_ie_i=0, e_1, e_1+e_2,  e_1 + e_2 + e_3$.

The embedding of $(\theta,n_ie_i)$ in the gauge degrees of freedom is
 represented
by an element $(\Theta, V)$, meaning boundary conditions
\begin{equation}
F(\sigma+1 ,t) = \Theta F(\sigma,t) + V + P
\label{Fbc}
\end{equation}
where $P \in \Lambda_{16}$ and $\Theta$ is an automorphism of $\Lambda_{16}$.
The full orbifold action is represented by elements $g=(g_{int}|g_{gauge})$
such as $(\theta,n_ie_i |\Theta, V)$.
In an orbifold compactification with Wilson line backgrounds \cite{inq},
the generators of the full orbifold action are $(\theta, 0 | 1, V)$ and
$(1, e_i | 1, a_i)$, where the $a_i$ are the Wilson lines.

In the permutation method we assign the operation $\Pi$ as a Wilson line in
the $e_1$ direction and rather consider the second generator to be
$(1, e_1 | \Pi, 0)$.  This then implies the following elements in the
simply-twisted sector
\begin{eqnarray}
g_0 &=& (\theta,  0 | 1, V)  \nonumber \\
g_1 &=&(\theta, e_1 | \Pi, \Pi V)  \nonumber \\
g_2 &=&(\theta, e_1 + e_2 | 1, \Pi V)  \nonumber \\
g_3 &=&(\theta, e_1+ e_2 + e_3 | \Pi, V)
\label{thsec}
\end{eqnarray}
Since $\theta$ is of order four and the $e_i$ are completely rotated by
 $\theta$,
$g_i^4 \equiv 1$ and $V$ must satisfy
\begin{equation}
2(V-\Pi V) \in \Lambda_{16}
\label{PiVcond}
\end{equation}
The interpretation of eqs.(\ref{thsec})
is that each $\theta$-subsector feels a different gauge action. This is
precisely what happens in an orbifold with quantized Wilson lines \cite{inq}.

For definiteness, in the following we will deal with the
$Spin(32)/Z_2$ lattice. The analysis can be straightforwardly
extended to the $E_8 \times E_8$ lattice.
We will consider a $\Pi$ action consisting of the permutation of two sets of
$L$
gauge bosonic coordinates. More precisely,
\begin{eqnarray}
&\Pi(F_1, \cdots, F_L, F_{L+1}, \cdots, F_{2L}, F_{2L+1}, \cdots, F_{16})
 \nonumber \\
&= (F_{L+1}, \cdots, F_{2L}, F_1, \cdots, F_L, F_{2L+1}, \cdots, F_{16})
\label{pidef}
\end{eqnarray}
In this $F$ basis, $\Pi$ acts as a non-diagonal matrix. It proves convenient to
work in another basis that diagonalizes $\Pi$. Then, for $J=1, \cdots, L$, we
define symmetric, $\ov{F}_J$, and antisymmetric, $\wh{F}_J$, combinations
\begin{eqnarray}
\ov{F_J} &= & \frac{F_J+ F_{J+L}}{\sqrt2}  \nonumber \\
\wh{F_J} &= & \frac{F_J-F_{J+L}}{\sqrt2}
\label{diagb}
\end{eqnarray}
Clearly, $\ov{F}_J \to \ov{F}_J$ and $\wh{F}_J \to  -\wh{F}_J$, under $\Pi$.

Given a 16-d vector $B$, we define its symmetric and antisymmetric parts as
\begin{eqnarray}
\ov{B} &=& (\frac{B_1 + B_{1+L}}{\sqrt2}, \cdots, \frac{B_L + B_{2L}}{\sqrt2} ;
      B_{2L+1}, \cdots, B_{16}) \nonumber \\
\wh{B} &=& (\frac{B_1 - B_{1+L}}{\sqrt2}, \cdots, \frac{B_L - B_{2L}}{\sqrt2} ;
      0, \cdots, 0)
\label{SAparts}
\end{eqnarray}
Notice that $B^2 = \ov{B}^2 + \wh{B}^2$.

For future purposes we also need to define $(16-L)$-dimensional lattices
$\Lambda_I$ and $\Lambda^*_I$, related to $\Lambda_{16}$.  A vector $P \in
\Lambda_{16}$ is of the form
\begin{eqnarray}
   P&=&(m_1 + \sh , \cdots, m_L + \sh, m_{L+1} + \sh , \cdots, m_{2L}+\sh,
m_{2L+1} + \sh , \cdots, m_{16}+\sh) \nonumber \\
s&=&0,1  \quad\quad ; \quad\quad  (m_1 + \cdots + m_{16}) = even
\label{PL}
\end{eqnarray}
The type of $\ov{P}$ constructed from a vector $P$ invariant under $\Pi$
motivates the definition of the invariant lattice $\Lambda_I$. Thus,  a vector
$P_I \in \Lambda_I$ is of the form
\begin{eqnarray}
   P_I&=&(\sqrt2[n_1 + \sh] , \cdots, \sqrt2[n_L + \sh];
   n_{2L+1} + \sh , \cdots, n_{16}+\sh) \nonumber \\
s&=&0,1  \quad\quad ; \quad\quad  (n_{2L+1} + \cdots + n_{16})  = even
\label{PLI}
\end{eqnarray}
The dual of $\Lambda_I$, denoted $\Lambda_I^*$, is the set of vectors
$\ov{Q}$ such that $P_I \cdot \ov{Q} =\,$ int. It is easy to see that
$\ov{Q} \in \Lambda_I^*$ must have the form
\begin{eqnarray}
   \ov{Q}&=&(\frac{n_1}{\sqrt2}, \cdots, \frac{n_L}{\sqrt2};
   n_{2L+1} + \sh , \cdots, n_{16}+\sh) \nonumber \\
s&=&0,1  \quad\quad ; \quad\quad  (n_1 + \cdots + n_L + n_{2L+1} +
\cdots + n_{16})  + sL = even
\label{QLD}
\end{eqnarray}

\subsection{The Partition Function}

For each element in the full orbifold group there are twisted sectors in which
 the
boundary conditions in $\sigma$ are of type (\ref{xbc}). Moreover, in each
 sector
there must be a projection on invariant states. This projection is implemented
by considering twisted boundary conditions in the $t$ direction. We will denote
by $(g;h)$ the boundary conditions in the $(\sigma;t)$ directions. It can be
shown \cite{orb2} that the partition function can be written as
\begin{equation}
Z = \frac{1}{|{\cal P}|} \sum_g \sum_{h | [h,g]=0} Z(g;h)
\label{pf}
\end{equation}
where $Z(g;h)$ is a path-integral evaluated with boundary conditions $(g;h)$.
$|{\cal P}|$ is the order of the orbifold point group.

The various pieces $Z(g;h)$ are related one another by modular
transformations of the world-sheet parameter $\tau$. More precisely, under
$\tau \rightarrow \frac{a \tau + b}{c \tau + d}$, the boundary conditions
change as $(g;h) \rightarrow (g^dh^c;g^bh^a)$. Therefore
\begin{equation}
Z(g;h) \rightarrow Z(g^dh^c; g^bh^a)
\label{Zmt}
\end{equation}
Hence, we can generate pieces of $Z$ by applying modular transformations.
Moreover, since in general $Z(g;h)$ picks up phases under these transformations
and there are some that leave the boundary conditions unchanged (e.g. $\tau
\rightarrow \tau + N$ in the $Z_N$ orbifold), there will appear modular
invariance restrictions on the possible forms of the full orbifold group.

$Z(g;h)$ is made up of contributions from the different string right- and
left-movers. The contribution from the space-time and internal bosonic
coordinates is well known \cite{imnq}, the left and right parts being
conjugate of each other since we are dealing with a symmetric orbifold.
The contribution from the right-handed fermions is given in terms of
$\vartheta$-functions and it depends only on the internal twist vector $v$.
For example, if $g_{int}=(\theta, n_ie_i)$, and $h_{int}=(1,0)$, we have up to
phases,
\begin{equation}
   Z_\psi(\theta, n_ie_i; 1,0) \sim  \sum_{\alpha,\beta=0,\oh} \,
 \ov{\eta}_{\alpha\beta}
   \prod_{a=0}^3  \,
   \frac{\vartheta \left[
   \begin{array}{c}
      {\alpha+v_a}\\
      {\beta}
 \end{array}
 \right ] (\bar \tau) }
 {\eta(\bar \tau) }
 \label{psiz}
 \end{equation}
where the sum is over spin structures. Modular invariance fixes
 $\ov{\eta}_{0\oh}=
\ov{\eta}_{\oh 0}=-\ov{\eta}_{00}$. We make the standard choices
 $\ov{\eta}_{00}=1$
and $\ov{\eta}_{\oh \oh}=1$. Using properties of $\vartheta$-functions,
 (\ref{psiz})
can be written as
\begin{equation}
  Z_\psi(\theta, n_ie_i; 1,0) \sim \frac{1}{\eta^4(\bar \tau)}  \! \sum_{
  \begin{array}{c}
     {r_a \in \IZ} \\
     {(r_0 + \cdots + r_3) = \hbox{odd}}
     \end{array}
     } \left [ \ov{q}^{\oh (r+v)^2} - \ov{q}^{\oh (r+v+S)^2} \right ]
     \label{psiz2}
\end{equation}
where $q=e^{2i \pi \tau}$, $S=(\oh,\oh,\oh,\oh)$ and here $v=(0,v_1,v_2,v_3)$.

The contribution of the gauge bosons in a $(g;h)$ sector can be computed as
$Tr (q^{H_F(g)} h)$. $H_F(g)$ is the Hamiltonian of the $F$ fields with
$\sigma$-boundary conditions twisted by $g$ and the trace is evaluated over
the corresponding Hilbert space. We will now review this construction when the
$g$ action on the $F$ is given by $(1,V)$. We will then extend the analysis to
embeddings of type $(\Pi,\Pi V)$.

We consider the general expansion
\begin{equation}
   F(\sigma,t) = F_0 + M \sigma_{-} + \frac{i}{2} \sum_r \,  \frac{\alpha_r}{r}
 \,
   e^{-2i\pi  r \sigma_{-}}
   \label{Fexp}
\end{equation}
where $\sigma_{-}=\sigma-t$. If  $g_{gauge} = g_V \equiv (1,V)$, the $\sigma$
boundary condition (\ref{Fbc}) implies that $M=P+V$ and that the oscillator
level $r$ is an integer. The Hamiltonian for the $F$ bosons is thus given by
\begin{equation}
H_F(g_V) = \oh (P+V)^2 + \sum_{n=1}^{\infty} \ :\alpha^{\dagger}_n \alpha_n :
- \frac{16}{24}
\label{FHal}
\end{equation}
Now, if the $h$ action on the $F$ is trivial, we have up to phases,
\begin{equation}
Z_F(g_V;1) = Tr(q^{H_F(g_V)}) \sim
\frac{1}{\eta^{16}(\tau)} \sum_{P \in \Lambda_{16}} \, q^{\oh (P+V)^2}
\label{Fz}
\end{equation}
Combining with (\ref{psiz2}) we obtain the complex piece of $Z(g;1)$ for
$g=(\theta,n_ie_i|1,V)$. For a $Z_4$ orbifold this piece must transform in
 itself
under $\tau \to \tau + 4$. From the transformations of (\ref{psiz2}) and
 (\ref{Fz})
we then obtain the modular invariance condition
\begin{equation}
   4(V^2 - v^2) = 0 \, mod \, 2
   \label{Vcond}
   \end{equation}
together with the embedding condition $4V \in \Lambda_{16}$.

For completeness we also consider the effect of a $t$ boundary condition $h$
encoded by the element $(1,V)$. This is
   \begin{equation}
   F(\sigma, t+1) = F(\sigma,t) + V + P'
   \label{Ftbc}
   \end{equation}
This tells us that in this case $h$ acts as a translation. Then, up to phases,
   \begin{equation}
      Z_F(g_V;g_V) = Tr(q^{H_F(g_V)} e^{2i\pi V\cdot M }) \sim
      \frac{1}{\eta^{16}(\tau)} \sum_{P \in \Lambda_{16}} \, q^{\oh (P+V)^2}
      e^{2i\pi (P+V)\cdot V} \label{Fz2}
      \end{equation}
If the phases in (\ref{psiz2}) and (\ref{Fz}) are chosen to be unity,
modular invariance implies an overall phase  $e^{-i\pi(V^2-v^2)}$ in
$Z_{\psi F}(g;g)$, $g=(\theta,n_ie_i|1,V)$  \cite{imnq}.

We now turn to the case of boundary conditions involving permutations.
We will consider $g_1 = (\theta,e_1|\Pi,\Pi V)$. Since $\Pi$ does not
affect the coordinates $F_K$, $K=2L+1, \cdots,16$; their expansions and
contribution to $Z_F$ are as before. Changes are due to the coordinates
$\ov{F}_J$ and $\wh{F}_J$, $J=1, \cdots, L$, with expansions
   \begin{eqnarray}
   \ov{F}(\sigma,t) &=& \ov{F}_0 + \ov{M} \sigma_{-} +
   \frac{i}{2} \sum_r \,  \frac{\ov{\alpha_r}}{r} \, e^{-2i\pi  r \sigma_{-}}
\nonumber \\
 \wh{F}(\sigma,t) &=& \wh{F}_0 + \wh{M} \sigma_{-} +
   \frac{i}{2} \sum_r \,  \frac{\wh{\alpha_r}}{r} \, e^{-2i\pi  r \sigma_{-}}
   \label{newFexp}
\end{eqnarray}
To simplify, in the above, the untouched coordinates coordinates $F_K$,
have been included together with the $\ov{F}_J$.
Since $g_{gauge}=g_\Pi \equiv (\Pi, \Pi V)$, the boundary conditions are
\begin{eqnarray}
 \ov{F}_J(\sigma+1,t) &=&  \ov{F}_J(\sigma,t) + \frac{P_J +V_{J+L} + P_{J+L} +
   V_J}{\sqrt2} \nonumber \\
 \wh{F}_J(\sigma+1,t) &=&  -\wh{F}_J(\sigma,t) + \frac{P_J +V_{J+L} - P_{J+L}
 - V_J}{\sqrt2} \nonumber \\
\ov{F}_K(\sigma+1,t) &=& \ov{F}_K(\sigma,t) + P_K + V_K
\label{newFbc}
\end{eqnarray}
Imposing these conditions to the expansions (\ref{newFexp}) implies that
the oscillators $\ov{\alpha}_r$ have integer levels $r=l$ whereas the
$\wh{\alpha}_r$ have $r=l + \oh$. Furthermore, the momenta must satisfy
\begin{eqnarray}
   \ov{M} &=& \ov{Q} + \ov{V} \nonumber \\
   \wh{M} &=& 0
   \label{Mcond}
   \end{eqnarray}
where $\ov{Q}$ and $\ov{V}$ are respectively the symmetric parts of
$P$ and $V$. In fact, $\ov{Q} \in \Lambda^*_I$, according to (\ref{QLD}).
Finally, the symmetric zero mode
$\ov{F}_{0J}$ is arbitrary while $2\wh{F}_{0J} = (P_J + V_{J+L} -
P_{J+L} - V_J)/\sqrt2$.

The Hamiltonian $H_F(g_\Pi)$ splits into a symmetric and antisymmetric
part, $H_F = \ov{H}_F + \wh{H}_F$, where
\begin{eqnarray}
\ov{H}_F(g_\Pi) &=& \oh (\ov{Q} + \ov{V})^2 + \sum_{n=1}^{\infty}
:\ov{\alpha}_n^{\dagger} \ov{\alpha}_n : -\frac{L}{24} -\frac{(16-2L)}{24}
\nonumber \\
\wh{H}_F(g_\Pi) &=& \sum_{r \in \IZ + \oh} :\wh{\alpha}_r^{\dagger}
\wh{\alpha}_r : + \frac{L}{48}
\label{HFs}
\end{eqnarray}
The $\sigma$ boundary conditions (\ref{newFbc}) then define states
$|\ov{M}, \ov{N}, \wh{N} \rangle$, where $\ov{N}$ and $\wh{N}$ are
oscillator numbers. Taking the trace over these states gives
\begin{equation}
Z_F(g_\Pi ; 1) = Tr(q^{H_F(g_\Pi)}) =
\sum _{{\bar Q} \in \Lambda _I^*}\frac { q^{ \frac{1}{2} (\ov Q + \ov V)^2 }}
{{\eta (\frac {\tau }2)}^L\eta(\tau)^{16-2L} }
\label{t1s}
\end{equation}
The $\eta(\tau)^{16-2L}$ in the denominator
arises from the $16 -2L$ coordinates untouched by
the permutation. The term ${\eta (\frac {\tau }2)}^{-L}$ originates as
\begin{equation}
{\eta (\frac {\tau }2)}^{-L} =
\left [ q^{ \frac {L}{24}}\prod _{n=1} {(1-q^n)}^L  \right ]^{-1}
\left [ q^{-\frac {L}{48}}\prod _{n=1} {(1-q^{n-1/2})}^L \right ]^{-1}
\label{etath}
\end{equation}
The first (second) factor comes from the trace over the $L$ symmetric
(antisymmetric) oscillators. Finally, in (\ref{t1s}), we have made a
definite choice of phases so as to have an equality.

Combining (\ref{t1s}) and (\ref{psiz2}) gives $Z_{\psi F}(g_1;1)$.
Performing a $\tau \to \tau + 4$ transformation
gives the modular invariance constraint
\begin{equation}
4\left [ \ov{V}^2 - v^2 + \frac{L}{8} \right ] = 0 \, mod \, 2
\label{barVcond}
\end{equation}
together with the embedding condition $4\ov{V} \in \Lambda_I$.

We next wish to generate other contributions to the partition function
by applying modular transformations to (\ref{t1s}). We begin with
$Z_F(1;g_{\Pi})$ which is obtained from a $\tau \to -1/\tau$ transformation
Using the Poisson resummation formula \cite{poisson} we obtain
\begin{equation}
Z_F(1;g_\Pi) =
\sum _{P_I \in \Lambda _I} \frac {
q^{\frac{1}{2} P_I^2 } e^{2i\pi P_I \cdot \ov{V}} }
{\eta (2\tau)^L \eta (\tau)^{16-2L} }
\label{1ts}
\end{equation}
We are also interested in deriving this result directly by imposing
$g_{\Pi}$ boundary conditions in the $t$ direction and computing
$Tr(q^{H_F(1)} g_\Pi )$. In this way we
will gain a better understanding about the action of the permutation $\Pi$.

The Hamiltonian $H_F(1)$ is essentially given in (\ref{FHal}) and
the trace must be taken over states $|P,N\rangle$. Now, to derive
the effect of inserting the $g_\Pi$ projection, it is convenient
to split $P$ into its $\ov{P}$ and $\wh{P}$ parts, as well as to
consider two set of oscillators $\ov{\alpha}_n$ and $\wh{\alpha}_n$.
For the antisymmetric momenta we have $\langle \wh{P} | \Pi \wh{P}
\rangle = 0$, unless $\wh{P}=0$. Therefore, recalling the definition
(\ref{SAparts}) of $\wh{P}$, we see that the only momentum states that
survive in the trace are those invariant under permutations, namely
those with $P_J = P_{J+L}$, for $J=1, \cdots, L$. For the symmetric
momenta, $\Pi | \ov{P} \rangle = | \ov{P} \rangle $ so that
$g_\Pi |\ov{P} \rangle = e^{2i\pi \ov{P} \cdot \ov{V}} | \ov{P} \rangle$.
Moreover, when $\wh{P}=0$, $\ov{P}_J= \sqrt2 P_J$, for $J=1, \cdots, L$.
Hence, $\ov{P} = P_I \in \Lambda_I$.

The above discussion explains the sum and the numerator in (\ref{1ts}).
The $\eta(2\tau)$ in the denominator appears because $\Pi$ multiplies
the $\wh{\alpha}_n$ oscillators by a phase $e^{i\pi}$. Indeed, the
permuted oscillator contribution is
\begin{equation}
  \left [ q^{ \frac {L}{24}}\prod _{n=1} {(1-q^n)}^L  \, \cdot \,
 q^{\frac {L}{24}}\prod _{n=1} {(1+q^n)}^L \right ]^{-1}
= \left [ q^{ \frac {2L}{24}}\prod _{n=1} {(1-q^{2n})}^L \right ]^{-1}
= \eta(2\tau)^{-L}
\label{etatt}
\end{equation}
Basically, we see that $\Pi$ also projects onto
invariant oscillator states.

We have now a neat interpretation of the action of $\Pi$ in the $t$
direction when inserted into the trace.
The factor of two in $\eta(2\tau)$ as well as the factor of $\sqrt2$
in $P_I$ can be heuristically explained as follows. Suppose that there
is a Hamiltonian $H=H_1+H_2$, where $H_1$ and $H_2$ have the same Hilbert
space. Suppose also that a permutation $\Pi$ exchanges states $|a_1 \rangle$
and $|a_2 \rangle$. Then,
\begin{equation}{}
\langle a_1 , a_2 | \Pi |a_1 , a_2\rangle  =
\left\{
\begin{array}{ll}
\langle a_1 , a_2 |a_1 , a_2\rangle  &
\quad \hbox{if} \quad  |a_1 \rangle = |a_2 \rangle\\
0  & \quad \hbox{otherwise}
\end{array}
\right.
\end{equation}
Therefore,
\begin{eqnarray}
Tr (q^{H_1+H_2} \Pi) &=&
\sum_{a_1,a_2} \langle a_1 , a_2 | q^{H_1+H_2} \Pi |a_1 , a_2\rangle
\nonumber \\
&=&  \sum_{a_1} \langle a_1 , a_1 | q^{H_1+H_2} |a_1 , a_1\rangle =
\sum_{a_1} \langle a_1 |q^{2H_1} | a_1 \rangle
\end{eqnarray}
This means that the two identical theories being permuted collapse
into one but evaluated at $2\tau$. This becomes even more explicit
if we rewrite eq. (\ref{1ts}) in terms of Riemann theta functions.

{}From $Z_F(g_\Pi ; 1)$ we can also generate $Z_F(g_\Pi^2; g_\Pi)$ thus
developing some insight about the doubly-twisted sector. The first
step is to apply a $\tau \to \tau + 2$ transformation to (\ref{t1s})
to arrive at
\begin{equation}
Z_F(g_\Pi; g_\Pi^2) =
e^{-2i\pi(\ov{V}^2 - \frac{L}{8})} \,
\sum _{{\ov{Q}} \in \Lambda _I^*}
\frac {
q^{ \frac{1}{2}  (\ov{Q} + \ov{V})^2 }
e^{2i\pi (\ov{Q} + \ov{V})\cdot 2\ov{V}} \, e^{2i\pi\ov{Q}^2} }
{{\eta (\frac {\tau }2)}^L \eta(\tau )^{16-2L} }
\label{t1t2s}
\end{equation}
where we have neglected constant phases that cancel against those in $Z_\psi$.
Since $\Lambda_I^*$ is not an integer lattice, here we cannot get rid of
the phase quadratic in $\ov{Q}$ as it is done in an ordinary level-one
$Z_F(g_V; g_V^2)$. This is an indication that in computing
(\ref{t1t2s}) directly as a trace, a $\ov{Q}$-dependent phase must be
allowed. This fact is not evident from the boundary conditions.

The next step is to derive $Z_F(g_\Pi^2; g_\Pi)$ by
applying to (\ref{t1t2s}) a $\tau \to -1/\tau$ transformation followed
by a $\tau \to \tau +1$ transformation. To this end it is
necessary to shift $\ov{Q}^2$ into a linear term in $\ov{Q}$ so that
the Poisson resummation formula can be used. In fact, from (\ref{QLD})
we see that
\begin{eqnarray}
\ov{Q}^2 & = &
\frac{n_1}2 + \cdots +\frac {n_L}2 + \frac {sL}2 \quad mod \quad 1 \nonumber \\
&=& 2\ov {Q} \cdot \ov{G} \quad  mod \quad 1
\label{Qs}
\end{eqnarray}
The auxiliary vector $\ov{G}$ is defined as
\begin{eqnarray}
2\ov {G} & \equiv & (\s2, \cdots , \s2 ; 0, \cdots, 0) +
\xi \Upsilon \nonumber \\
\Upsilon  & \equiv & (0, \cdots, 0 ; 1, 0, \cdots , 0)
\label{Gdef}
\end{eqnarray}
with $\xi=0$ for $L$ even and $\xi=1$ for $L$ odd. Hence,
$e^{2i\pi\ov{Q}^2} = e^{2i\pi\ov{Q} \cdot 2\ov{G}} $.

After using the Poisson resummation formula we arrive finally at
\begin{equation}
Z_F(g_\Pi^2; g_\Pi) =
e^{-2i\pi(\ov{V}^2 - 2\ov{G}^2 - \frac{L}{8})} \,
\sum _{P_I \in \Lambda _I}
\frac {
q^{ \frac{1}{2}  (P_I + 2\ov{V} + 2\ov{G})^2 }
e^{2i\pi (P_I + 2\ov{V} + 2\ov{G})\cdot \ov{V}} }
{\eta (2\tau)^L \eta(\tau )^{16-2L} }
\label{t2t1s}
\end{equation}
{}From the above we see that to a $g_\Pi$-twisted sector there
corresponds a $g_\Pi^2$-twisted sector in which the allowed momenta
that survive the $g_\Pi$ projection are of the form $P_I + 2\ov{V} + 2\ov{G}$.
We now turn to deriving this result from boundary conditions in order
to better understand the meaning of the vector $2\ov{G}$.

Since $\Pi$ is of order two, we assume that $g_\Pi^2 = (1;2V) = g_{2V}$.
Then, in the $g_\Pi^2$-twisted sector, the Hamiltonian is of the form
(\ref{FHal}) and
\begin{equation}
Z_F(g_\Pi^2;g_\Pi) = Tr(q^{H_F(g_{2V})} g_\Pi)
\label{Fz21}
\end{equation}
The trace is taken over states $|M, N \rangle$, where $M=P+2V$. Now, as
we have seen before, inserting $g_\Pi$ in the trace has the effect of
projecting onto invariant oscillator states and this is reflected in
the $\eta (2\tau)^L$ factor. Also, $g_\Pi$ projects onto invariant
momenta, selecting states with $\wh{M} = 0$. This implies
\begin{equation}
P_J + 2V_J = P_{J+L} + 2V_{J+L} \quad ; \quad J=1, \cdots, L
\label{PI1}
\end{equation}
The shift $V$ must satisfy $2(V_J - V_{J+L})= \,$int, since
for any $P \in \Lambda_{16}$, $P_I-P_K =\,$ int.
This is guaranteed by the
embedding condition (\ref{PiVcond}). Then, using (\ref{PI1})
we can assert that the lattice momenta that survive are such that
\begin{eqnarray}
M&=&(m_1 + \sh , \cdots, m_L + \sh, m_1 + \sh , \cdots, m_{L}+\sh ;
m_{2L+1} + \sh , \cdots, m_{16}+\sh)
\nonumber \\
&+& 2(V_1, \cdots , V_L, V_1, \cdots V_L ; V_{2L+1}, \cdots V_{16})
\label{Minv1}
\end{eqnarray}
Clearly, $\wh{M} =0$.

To compute $\ov{M}$, we first notice that, since
$P=(M-2V) \in \Lambda_{16}$, it must be that
\begin{equation}
{\cal S} + (m_{2L+1} + \cdots + m_{16}) = \, even
\label{Smcond}
\end{equation}
where
\begin{equation}
{\cal S} = \sum_{J=1}^{L} \, 2(V_J - V_{J+L})
\label{Sdef}
\end{equation}
Thus, for ${\cal S} =$even, $(m_{2L+1} + \cdots + m_{16})=$even. To
have the same type of constraint when ${\cal S} =$odd, we redefine
$m_{2L+1} \to m_{2L+1} + 1$. With this convention it then follows
that $\ov{M}$ can be written as
\begin{equation}
\ov{M} = P_I + 2\ov{V} + 2\wh{V} + \zeta \Upsilon
\label {Minv2}
\end{equation}
where $P_I \in \Lambda_I$ and $\zeta = {\cal S} \,$mod 2.

The upshot of the above discussion is that computing directly from
boundary conditions leads to $Z_F(g_\Pi^2; g_\Pi)$ of the form
(\ref{t2t1s}), provided that
\begin{equation}
2\wh{V} + \zeta \Upsilon = 2\ov{G} \quad  mod \quad \Lambda_I
\label{VGrel}
\end{equation}
Moreover, in this case
\begin{equation}
2\ov{G}^2 + \frac{L}{8} = \wh{V}^2 \quad mod \quad 1
\label{G2cond}
\end{equation}
Hence, the overall phase in (\ref{t2t1s}) agrees with the expected
value $e^{-2i\pi V^2}$.

It is interesting to notice that, formally, the above results look quite
similar to those obtained in refs.\,\cite{fuchs,alda} (in the context of
$N=2$ coset model contructions) where characters $\chi(M\tau)$
emerge after modding
out by a cyclic permutation symmetry of $M$ internal theories.
In fact, the origin
of the factor $M$ is the same, the collapse of $M$ internal theories.
However, there are two relevant differences. On the one hand, since we are
considering permutations
in the gauge sector, this operation is intrinsically asymmetric in the
sense that it affects only the left sector. As a consequence, stringent
constraints such as eqns. (\ref{barVcond}) and (\ref{VGrel})
must be satisfied in order to match left and right sectors while
level matching is automatic when permuting similar theories in both sectors.
On the other hand, let us stress that we are introducing permutations here
as Wilson lines, associated
to given directions in the internal lattice.
The underlying model here is not necessarily invariant
under permutations.
Of course, if a model invariant under permutation of gauge factors,
in the sense that if a
representation $(R,R')$ appears so does $(R',R)$, is
constructed, then we can mod out by this symmetry.
This modding is an external projection unrelated to orbifold
operations. The addition of sectors twisted by permutations
to recover modular invariance,
will lead to even more severe contraints, because now $\Pi$ will
be felt in all $\theta ^n$ sectors. For example, in the $Z_4$
orbifold case, only permutations of four or
eight gauge bosonic coordinates are allowed.

The previous analysis has focused on the $Z_4$ orbifold. We will now
consider the $Z_2 \times Z_2$ orbifold with $SO(4)^3$ lattice. In this
case there are two internal twist generators $\theta$ and $\omega$ with
twist vectors $a=(0,1/2,0,-1/2)$ and $b=(0,0,1/2,-1/2)$. In the absence
of Wilson lines the generators of the full orbifold group are
$(\theta,0 | 1, A)$ and $(\omega,0 | 1, B)$, where $2A \in \Lambda_{16}$
and $2B \in \Lambda_{16}$. The permutation $\Pi$ is
embedded as a Wilson line in the first $SO(4)$ sub-lattice, the
corresponding generator being $(1,e_1 | \Pi, 0)$.

As reviewed in ref.\,\cite{afiu}, there are several ways in which the
twists $\omega$ and $\theta$ can act on the six dimensional lattice. They
have different fixed points structure and therefore lead to different
multiplicities in the spectrum.
Moreover, consistency constraints, due to the inclusion of Wilson lines,
also depend on the realization.

Let us consider the realization
\begin{equation}
\theta=(-1,1,-1) \quad ; \quad \omega=(1,-1,-1)
\label{mult16}
\end{equation}
Since $\omega$ does
not rotate the first sub-lattice, the $\omega$ sector is not split by
$\Pi$. On the other hand, the $\theta$ sector is split in a way encoded
by the elements $(\theta,0 | 1, A)$, $(\theta,e_1 | \Pi, \Pi A)$,
$(\theta,e_2 | \Pi, A)$ and $(\theta,e_1 + e_2 | 1, \Pi A)$. Similarly, the
relevant elements in the $\theta\omega$ sector are $(\theta\omega,0 | 1, C)$,
$(\theta\omega,e_1 | \Pi, \Pi C)$, etc., where $C=A-B$.

Defining $g_{\Pi A}= (\Pi, \Pi A)$, we find that the partition function
$Z_F(g_{\Pi A};1)$ is of the form (\ref{t1s}). Combining with the fermionic
piece and performing a $\tau \to \tau + 2$ transformation, that must leave
the partition function invariant, we find the constraints
\begin{eqnarray}
\ov{A}^2 - a^2 + \frac{L}{8}  &=& 0 \, mod \, 1  \nonumber \\
2\ov{A} &=& 2\ov{G} \, mod \, \Lambda_I
\label{barAcond}
\end{eqnarray}
Similar conditions must hold for $\ov{C}$. It then follows that
\begin{equation}
2\ov{B} \in \Lambda_I
\label{barBcond}
\end{equation}
The constraints (\ref{barAcond}) and (\ref{barBcond}) supplement the
regular modular invariance conditions $(A^2-a^2)=\,$mod 1,
$(B^2-b^2)=\,$mod 1 and $(A\cdot B-a\cdot b)=\,$mod 1.

The partition function  $Z_F(g_{\Pi A}; g_B)$ cannot be connected to
$Z_F(g_{\Pi A};1)$ by modular transformations. We then choose to connect
it to $Z_F(g_B;g_{\Pi A})$ which in turn is computed as
$Tr(q^{H_F(g_B)} g_{\Pi A})$ using our prescription for the action of
$g_{\Pi A}$ when inserted into the trace. With the usual choice of
phase we then find
\begin{equation}
Z_F(g_B; g_{\Pi A}) =
e^{-i\pi A \cdot B} \,
\sum _{P_I \in \Lambda _I}
\frac {
q^{ \frac{1}{2}  (P_I + B_I)^2 }
e^{2i\pi (P_I + B_I)\cdot \ov{A}} }
{\eta (2\tau)^L \eta(\tau )^{16-2L} }
\label{z2t2t1s}
\end{equation}
The shift $B_I$ appears when projecting into states with $\wh{M}=0$,
where $M=P+B$ and it is given by
\begin{equation}
B_I = \ov{B} + \wh{B} + \zeta_B \Upsilon
\label{Binv}
\end{equation}
where $\zeta_B = {\cal S}_B\,$mod 2 and
${\cal S}_B = \sum_{J=1}^{L} \, (B_J - B_{J+L})$. In fact, the existence
of solutions to $\wh{M}=0$ requires
\begin{equation}
B_J - B_{J+L} = 0 \, mod \, 1
\label{Bcond}
\end{equation}
Equivalently, $(\wh{B} + \zeta_B \Upsilon) \in \Lambda_I^*$. A simple way
to satisfy the constraints on the shift $B$ is to choose $\wh{B}=0$. In
this case we find
\begin{equation}
Z_F(g_{\Pi A}; g_B) =
e^{-i\pi \ov{A} \cdot \ov{B}} \,
\sum _{{\ov{Q}} \in \Lambda _I^*}
\frac {
q^{ \frac{1}{2}  (\ov{Q} + \ov{A})^2 }
e^{2i\pi (\ov{Q} + \ov{A})\cdot \ov{B}} }
{{\eta (\frac {\tau }2)}^L \eta(\tau )^{16-2L} }
\label{tatb}
\end{equation}
where we have used that $B_I = \ov{B}$.

An alternative to the above realization of the twists, usually more
attractive
since it leads to lower multiplicities, corresponds to the choice
\begin{equation}
\theta =  (-1, \sigma_1, \sigma_1) \quad ; \quad
\omega = (-\sigma_1, -1, -\sigma_1)
\label{mult31}
\end{equation}
In this case the fixed sets in the $\theta$ sector that feel the
action of $\Pi$ are not fixed by either $\omega$ or
$\theta \omega $. As a consequence, contributions such as
$Z_F(g_B; g_{\Pi A})$ or $Z_F(g_{\Pi A}; g_B)$ will not appear.
Therefore, no extra constraints apart from eq. (\ref {barAcond}) will be
required.

To end this section let us briefly consider the contribution of the
internal bosonic coordinates. For our purposes it is sufficient to know
that
\begin{equation}
Z_B(g;h) = \tilde{\chi}(g,h)
\left \{ \ov{q}^{E_0(g)-\frac{6}{24}} (1 + \cdots) \right \}
\left \{ q^{E_0(g)-\frac{6}{24}} (1 + \cdots) \right \}
\label{ZB}
\end{equation}
where the ellipsis stands for higher powers of $q$ corresponding to
oscillator states. The twisted vacuum energy $E_0$ depends on the
$g_{int}$-twist vector $(v_1,v_2,v_3)$ according to
\begin{equation}
E_0(g) = \sum_{a=1}^3 \, \oh |v_a|(1-|v_a|)
\label{Evac}
\end{equation}
$\tilde{\chi}(g,h)$ is a numerical factor that basically counts the number
of simultaneous fixed points of $g_{int}$ and $h_{int}$. In the
presence of Wilson lines it is convenient to decompose it as a sum of
terms $\tilde{\chi}(g,h|x_g)$ where the $x_g$ are the fixed points of
$g_{int}$. For more details we refer the reader to the Appendix of \cite{afiu}.

\subsection{Massless Spectrum and Generalized GSO Projectors}

In the last section we have seen how modular invariance and the operator
interpretation of the partition function impose constraints on the shift
$V$ and the number of permuted coordinates. We now want to discuss how
the partition function determines the allowed massless states.

The massless spectrum can be divided into twisted sectors acoording to
the $\sigma$ boundary conditions. Let us begin with the simply-twisted
sector, which in turn is divided into four sub-sectors according to
(\ref{thsec}). The partition function in each sub-sector is obtained
from $Z(g_i;1)$ by applying $\tau \to \tau + 1$ transformations. Moreover,
since $\forall P \in \Lambda_{16}$, $\Pi P \in \Lambda_{16}$ and
$\ov{(\Pi V)} = \ov{V}$, it can be shown that $Z(g_0;1) = Z_(g_2;1)$
and $Z(g_1;1) = Z_(g_3;1)$. This implies that the simply-twisted partition
function actually splits into two pieces that we will denote $Z(\theta)$
and $Z(\theta_\Pi)$. Explicitly,
\begin{eqnarray}
Z(\theta) &=& \frac{1}{2} \left [ Z(g_0;1) + Z(g_0;g_0) + Z(g_0;g_0^2) +
Z(g_0;g_0^3) \right ] \nonumber \\
Z(\theta_\Pi) &=& \frac{1}{2} \left [ Z(g_1;1) + Z(g_1;g_1) + Z(g_1;g_1^2) +
Z(g_1;g_1^3) \right ]
\label{Zth}
\end{eqnarray}
The $Z(g_i;g_i^n)$ only differ in phase factors, the powers of $q, \ov{q}$
are identical and fix the mass level. Actually, for massless states the
phase factors are all unity. The left masslessness conditions are
\begin{eqnarray}
Z(\theta) &:&  M_L^2 = N_L + \frac{1}{2} (P+V)^2 +
E_0(\theta) -1 = 0 \nonumber \\
Z(\theta_\Pi) &:&  M_L^2 = N_L + \frac{1}{2} (\ov{Q}+ \ov{V})^2 +
E_0(\theta) + \frac{L}{16} -1  = 0
\label{thLmass}
\end{eqnarray}
where $N_L$ includes the $x$- and $F$-oscillator numbers. The right
masslessness condition, common for both sectors, is given by
\begin{equation}
M_R^2 = N_R + \frac{1}{2} (r+v)^2 + E_0(\theta) -\oh = 0
\label{thRmass}
\end{equation}
States satisfying (\ref{thLmass}) and (\ref{thRmass}) appear with multiplicity
$2\tilde{\chi}(g_i,1|x_{g_i})$. This is multiplicity 8
in the $Z_4$ orbifold with $SU(4) \times SU(4)$ lattice.

Let us now consider the doubly-twisted sector. Since $g_\Pi^2 = g_V^2 =
g_{2V}$,
the partition function in this sector cannot be split into sub-sectors. Rather,
there is just a partition function $Z(\theta^2)$ and all massless states
satisfy
\begin{eqnarray}
Z(\theta^2) &:&  M_L^2 = N_L + \frac{1}{2} (P+2V)^2 +
E_0(\theta^2) -1 = 0 \nonumber \\
& &  M_R^2 = N_R + \frac{1}{2} (r+2v)^2 + E_0(\theta^2) -\oh =0
\label{th2mass}
\end{eqnarray}
$Z(\theta^2)$ includes terms $Z(g_i^2;1)$ and their $\tau \to \tau +1$ modular
transforms $Z(g_i^2;g_i^2)$, which only differ in phases that actually
vanish for massless states. There are also terms $Z(g_i^2;g_i)$ and their
$\tau \to \tau +1$ modular transforms $Z(g_i^2;g_i^3)$, that also contribute
equally to massless states. Moreover, they are obtained from modular
transformations applied to $Z(g_i;g_i^2)$ as explained before. In particular,
from (\ref{t2t1s}) we see that $Z(g_1^2;g_1)$ can only contribute to invariant
gauge momentum and oscillator states. The conclusion is that there are
two types of massless states according to whether or not they are invariant
under $\Pi$. In the $Z_4$ orbifold with $SU(4) \times SU(4)$ lattice,
their respective multiplicities turn out to be
\begin{eqnarray}
D_{inv}(\theta^2) &=& 2 + 2\, e^{2i\pi\Delta_2} \nonumber \\
D_{non-inv}(\theta^2) &=& 2 + e^{2i\pi\Delta_2}
\label{th2mult}
\end{eqnarray}
The phase $\Delta_n$ is given by
\begin{equation}
\Delta_n = (P+nV) \cdot V - (r+nv) \cdot v - \frac{n}{2} (V^2 - v^2) - \rho
\label{delph}
\end{equation}
where $\rho$ appears in the case of $x$-oscillator states and depends on
how the oscillator is rotated by $\theta$.

Finally, let us discuss the untwisted sector. All massless states satisfy
\begin{eqnarray}
Z(1) &:&  M_L^2 = N_L + \frac{1}{2} P^2 - 1 = 0 \nonumber \\
& &  M_R^2 = N_R + \frac{1}{2} r^2 -\oh =0
\label{1mass}
\end{eqnarray}
The surviving states must fulfill the usual projection $\Delta_0 =\,$int.
We must also construct $\Pi$-invariant combinations since we are
embedding $\Pi$ as a Wilson line. This is the equivalent of the condition
$P \cdot a_i = 0$ on massless states in the presence of quantized Wilson
lines.

In the case of $Z_2 \times Z_2$ with twists (\ref{mult16}),
the partition function in the $\theta$
sector also divides into two distinct pieces originating sub-sectors
$\theta$ and $\theta_\Pi$. The $\theta\omega$ sector splits in a similar
way. The $\omega$ sector is no split but invariant and non-invariant states
are projected differently. In the alternative case (\ref{mult31}),
the $\theta$ sector also divides into two sub-sectors. The $\omega$
and the $\theta\omega$ sectors are unaffected.
The relevant mass formulae and multiplicities
follow from the form of the partition function discussed in the previous
section.

\section{Examples of $SO(10)$ and $SU(5)$ level 2 models}

In this section we apply the techniques developed in the
previous section to the construction of
explicit GUT examples.
We present a couple of four-generation $SO(10)$ models
with an adjoint Higgs in the massless spectrum
(we were only able to construct models with a $54$
and no adjoint in ref.\cite{afiu} ). We also
build a four-generation $SU(5)$ model.

A natural extension of the permutation method just described
is to combine the action of $\Pi$ with a shift $S$ in $\Lambda_{16}$.
More precisely, we consider a generator $(1, e_1 | \Pi, S)$. Since $\Pi$ is
of order two, it must be that $2S \in \Lambda_{16}$. In the $Z_4$ orbifold,
the simply-twisted sector splits according to the elements
$(\theta, 0 | 1, V)$, $(\theta, e_1 | \Pi, S + \Pi V)$,
$(\theta, e_1 + e_2 | 1, S + \Pi S + \Pi V)$ and
$(\theta, e_1+ e_2 + e_3 | \Pi, \Pi S + V)$. The original translation $V$
must satisfy eq. (\ref{PiVcond}).
The structure of the partition function is completely analogous to the case
$S=0$. In particular, there appear modular invariant conditions of the
form of eqs. (\ref{barVcond}) and (\ref{VGrel}), with $V$ replaced by $V+S$.

The main effect of the shift $S$ is to change the projection onto invariant
states in the untwisted sector. States with invariant momentum, $\Pi P = P$,
must satisfy $P.S=\,$int. On the other hand, if $\Pi P \not= P$,
$P\cdot (S - \Pi S)=\,$int and $P\cdot (V - \Pi V)=\,$int,
the states survive into combinations invariant under the full action of
$\Pi$ and $S$. This means, for instance, that when $P\cdot S = \oh$,
the combination changes sign under $\Pi$. As a consequence, antisymmetric
representations such as adjoints in $SO(10)$, can arise in the untwisted
sector. This result will be illustrated in our first example.

\subsection{\bf $SO(10)_2$ model with an adjoint and 4 generations.}

This model is based on the $Z_4$ orbifold with $SU(4)\times SU(4)$
compactifying lattice. The starting $SO(10)\times SO(18)\times SU(2)\times
U(1)$ model is derived by a gauge embedding with shift
\begin{equation}
V=\frac 14 (2,2,2,2,2,\, 0,0,0,0,0,\, 0,0,0,0,1,1)
\end{equation}
An $SO(10)$ gauge group at level 2 emerges
after implementing a permutation $\Pi$ of the form (\ref{pidef}) with $L=5$,
accompanied by a shift
\begin{equation}
\label{s45}S=\frac 14(1,1,1,1,1,\, 1,1,1,1,1,\, 1,1,1,1,3,3)
\end{equation}
The resulting model has gauge group $SO(10)\times SU(4)\times SU(2)\times
U(1)^2$ where $SO(10)$ is realized at level two.
The massless spectrum is found basically as described in section 2.3.
There are 8 generations. More notably, there is an adjoint $45$ in the
$U_3$ untwisted sector. Let us now explain how this $45$ does materialize.

\begin{table}
\begin{center}
\begin{tabular}{|c|c|c|c|c|}
\hline
$Sector$
   & $SO(10)\times SU(2 )^3 \times U(1)^3 \hspace {.1cm}  $ & $Q_1$ & $Q_2$ &
 $Q_3$  \\
\hline
$  U_1, U_2 $   &    2 (1,1,2,2)  &  0  &   $1/2$   &   $ 1/2 $  \\
\hline
$  U_3   $      &      (45,1,1,1)           &  0     &   0   &   0  \\
\hline
\hline
$ (\theta,V)$      &    4 ($\ov {16}$,1,1,1)     & 0     &  0 &  $  1/4 $  \\
\hline
\hline
$ (\theta, V+W)$      &   no massless states     &      &  &   \\
\hline
\hline
$(\theta ,\ov{V}+\ov{W})$                &    4 (10,1,1,1)     & $  1/4 $
&
 $  1/4 $  &   0  \\
\hline
                &    4 (1,2,1,1)     & $  -1/4 $      &$  1/4 $  &0   \\
\hline
           &    4 (1,1,2,1)     & $  +1/4 $      &$  -1/4 $  &0   \\
\hline
                &    4 (1,1,1,1)     & $  -1/4 $      & $  -1/4 $  & $  1/2 $
\\
\hline
                &    4 (1,1,1,1)     & $  -1/4 $      & $  -1/4 $  & $  -1/2 $
 \\
\hline
\hline
$(\theta, \ov{V}+ \ov{S}+ \ov{W})$     &    4 (10,1,1,1)     & $  -1 /4 $
      & $  1/4 $  &   0  \\
\hline
                &    4 (1,2,1,1)     & $  1/4 $      &$  1/4 $  &0   \\
\hline
                &    4 (1,1,2,1)     & $  -1/4 $      &$  -1/4 $  &0   \\
\hline
                &    4 (1,1,1,1)     & $ 1/4 $      & $  -1/4 $  & $  1/2 $  \\
\hline
                &    4 (1,1,1,1)     & $  1/4 $      & $  -1/4 $  &$  -1/2 $ \\
\hline
\hline
$(\theta ^2,2V)$           &    4 (10,1,1,1)       & 0     & 0 &   $  -1/2 $ \\
\hline
           &    4 (10,1,1,1)       & 0     & 0 &   $  1/2 $ \\
\hline
           &    3 (1,2,1,1)       & $  1/2 $ & 0   &   $  1/2 $ \\
\hline
           &    1 (1,2,1,1)       & $  -1/2$  & 0  &   $  -1/2 $  \\
\hline
           &    3 (1,2,1,1)       & $  -1/2 $& 0  &   $  1/2  $\\
\hline

           &    1 (1,2,1,1)       & $  1/2$  & 0  &   $  -1/2 $  \\
\hline

       &    3 (1,1,2,1)       & 0     & $  1/2$   &   $  1/2 $  \\
\hline
           &    1 (1,1,2,1)       & 0     & $  -1/2$   &   $  -1/2 $  \\
\hline
           &    3 (1,1,2,1)       & 0     & $  1/2  $ &   $  -1/2 $  \\
\hline
           &    1 (1,1,2,1)       & 0     & $  -1/2  $ &   $  1/2 $  \\
\hline
$Osc.$     &   2$\times $ 3(1,1,1,2)         & 0     & 0  &   0   \\

          &   2$\times $ 1 (1,1,1,2)         & 0     & 0  &   0   \\
\hline
\end{tabular}
\end{center}
\caption{Particle content of the model of section 3.1.}
\label{t31}
\end{table}
\bigskip

In the $U_3$ sector, states must have $P\cdot V = \pm \oh$. In the original
model there are states with
\begin{equation}
P_{10,10} = (\underline{\pm 1,0,0,0,0},\, \underline{\pm 1,0,0,0,0},\,
0,0,0,0,0,0)
\label{pten}
\end{equation}
where underlining means permutations. These states are part
of a $(10,18)$ of $SO(10)\times SO(18)$. Some of the above momenta
are invariant under $\Pi$ and are projected out because they have
$P_{10,10} \cdot S = \pm \oh$. There are also non-invariant momenta with
$P_{10,10} \cdot S = 0$ that give rise to invariant combinations such as
\begin{equation}
\left [ (+1,0,0,0,0,\, -1,0,0,0,0,\, 0,\cdots, 0) \oplus
(-1,0,0,0,0,\, +1,0,0,0,0,\, 0,\cdots, 0) \right ]
\label{t25}
\end{equation}
Finally, there are non-invariant momenta with $P_{10,10} \cdot S = \pm \oh$
that give rise to invariant combinations of the form
\begin{equation}
\pm \left [ (+1,0,0,0,0, \, 0,+1,0,0,0,\, 0,\cdots, 0) \ominus
(0,+1,0,0,0,\, +1,0,0,0,0, \, 0,\cdots, 0) \right ]
\label{t20}
\end{equation}
The $45$ comprises all 25 states of type (\ref{t25}) plus all 20 states of
type (\ref{t20}).

The number of generations may be further reduced by associating a Wilson line
$W$ ($W=\Pi W$) to the second $SU(4)$ lattice, for instance
\begin{equation}
W=\frac 14(0,0,0,0,0, \, 0,0,0,0,0,\, 2,2,0,0,0,0)
\label{4gen}
\end{equation}
This second order Wilson line breaks the $SU(4)\times U(1)$ group above into
$SU(2)\times SU(2)\times U(1)^2$ and reduces the number of $SO(10)$
generations to 4. The massless spectrum of the model is given
in Table \ref{t31}.
\subsection{\bf $SU(5)_2$ model with 4 generations.}

This model is straightforwardly obtained by interchanging the roles of $S$ and
$W$ in the above $SO(10)_2$ example. In this way, a
$SU(5)_2\times SU(2)^3\times U(1)^4$ model with four generations finally
emerges. The massless spectrum of the model is shown in Table \ref{t32}.
\begin{table}
\begin{center}
\begin{tabular}{|c|c|c|c|c|c|}
\hline
$Sector$
   & $SU(5)_2\times SU(2 )^3 \times U(1)^4 \hspace {.1cm}  $ & $Q_0$ & $Q_1$ &
 $Q_2$ & $Q_3$  \\
\hline
$  U_1, U_2 $   &    2 (1,1,1,2)  & 0 &  0  &   $1/2$   &   $ 1/2 $  \\
\hline
$  U_3   $      &      (24+1,1,1,1) & 0           &  0     &   0   &   0  \\
\hline
\hline
$ (\theta,V)  $      &    4 ($\ov{10}$,1,1,1)  &$1/2$   & 0     &  0 &  $  1/4
 $  \\
\hline
               &    4 (5 ,1,1,1)  & $-3/2$   & 0     &  0 &  $  1/4 $  \\
\hline
                &    4 (1,1,1,1)  & $5/2$   & 0     &  0 &  $  1/4 $  \\
\hline
\hline
$(\theta,V+W)$                &    4 (5,1,1,1)  & 1   & $  1 /4 $      & $  1/4
 $  &   0  \\
\hline
                &    4 $(\ov{5},1,1,1)$  & $-1$   & $  1 /4 $      & $  1/4 $
 &   0  \\
\hline
           &    4 (1,2,1,1)  & 0    & $  -1/4 $      &$  1/4 $  &0   \\
\hline
                &    4 (1,1,2,1)     & 0 & $  1/4 $      &$  -1/4 $  &0   \\
\hline
                &    4 (1,1,1,1)     & 0 & $  -1/4 $      & $  -1/4 $  & $  1/2
 $   \\
\hline
                &    4 (1,1,1,1)     & 0 & $  -1/4 $      & $  -1/4 $  & $
-1/2
 $  \\
\hline
\hline
$(\theta,\ov{V}+\ov{S})$       &  no massless states     & &       &  &
 \\
\hline
\hline
$(\theta,\ov{V}+\ov{S}+\ov{W})$                &    4 (5,1,1,1)  & $1$   &
  $  -1 /4 $      & $  1/4 $  &   0  \\
\hline
                &    4 $(\ov{5},1,1,1)$  & $-1  $ &  $  -1 /4 $      & $  1/4
 $  &   0  \\
\hline
             &    4 (1,2,1,1)     & 0 & $  1/4 $      &$  1/4 $  &0   \\
\hline
                &    4 (1,1,2,1)     & 0 & $  -1/4 $      &$  -1/4 $  &0   \\
\hline
                &    4 (1,1,1,1)     & 0 & $  1/4 $      & $  -1/4 $  & $  1/2
$
   \\
\hline
                &    4 (1,1,1,1)     & 0 & $   1/4 $      & $  -1/4 $  & $
-1/2
 $  \\
\hline
\hline
$\theta ^2$
           &    4 (5,1,1,1)  & 1      & 0     & 0 &   $  1/2 $ \\
\hline
           &   $ 4 (\ov{5},1,1,1)$  & $-1$      & 0     & 0 &   $  1/2 $ \\
\hline
           &    4 (5,1,1,1)  & 1      & 0     & 0 &   $  -1/2 $ \\
\hline
           &   $ 4 (\ov{5},1,1,1)$  & $-1$      & 0     & 0 &   $  -1/2 $ \\
\hline
           &    3 (1,1,2,1)  & 0     & 0     & $  1/2$   &   $  1/2 $  \\
\hline
           &    1 (1,1,2,1)     & 0  & 0     & $  -1/2$   &   $  -1/2 $  \\
\hline
           &    3 (1,1,2,1)     & 0     & 0     & $  1/2  $ &   $  -1/2 $  \\
\hline
           &    1 (1,1,2,1)  & 0     & 0     & $  -1/2  $ &   $  1/2 $  \\
 \hline

           &    3 (1,2,1,1)   & 0    & $  1/2 $ & 0   &   $  1/2 $ \\
\hline
           &    1 (1,2,1,1)   & 0    & $  -1/2$  & 0  &   $  -1/2 $  \\
\hline
           &    3 (1,2,1,1)   & 0 & $ - 1/2 $& 0  &   $   1/2  $ \\
\hline
           &    1 (1,2,1,1)   & 0 & $  1/2 $& 0  &   $   -1/2  $ \\
 \hline

$Osc.$      &   2$\times $ 3(1,1,1,2)  & 0       & 0     & 0  &   0   \\
\hline
            &   2$\times $ 1(1,1,1,2)  & 0       & 0     & 0  &   0   \\ \hline
\end{tabular}
\end{center}
\caption{Particle spectrum of the model of section 3.2.}
\label{t32}
\end{table}
Interestingly enough, the two examples above show two possible uses of the
inclusion of a shift accompanying the permutation twist. In the $SO(10)$
case it helps in picking up the adjoint (asymmetric) representation in the
decomposition $(10,10)=54+1+45$, in the $SU(5)$ case it reduces the number
of generations.

\subsection{\bf $SO(10)_2$ model with an
adjoint and a $(16 + \ov{16})$ pair}

This example is based on $Z_2 \times Z_2$ with the twist realization
(\ref{mult31}). The starting $SO(10)^3 \times U(1)$ model is derived
with the gauge embedding
\begin{eqnarray}
A &=& \oh (1,1,1,1,1, \, 0,0,0,0,0, \, 1,0,0,0,0,0) \nonumber \\
B &=& \oh (0,0,0,0,0, \, 0,0,0,0,0, \, 1,1,1,1,1,-1)
\label{ex3}
\end{eqnarray}
To obtain $SO(10)$ at level two we act with a permutation $\Pi$ of
type (\ref{pidef}) with $L=5$, plus a shift
\begin{equation}
S=\frac 14(1,1,1,1,1,\, 1,1,1,1,1,\, 1,1,1,1,1,-3)
\label{Sz2z2}
\end{equation}
The resulting model has gauge group $SO(10)\times SU(5)\times
U(1)^2$, with $SO(10)$ realized at level two. The massless
spectrum, shown in Table \ref{t33}, is found along the usual procedure.
There are 4 net generations and a $(16 + \ov{16})$ pair.
There is an adjoint $45$ in the $U_1$ untwisted sector that appears
exactly as explained in section 3.1. The normalization of the $U(1)$
charges is chosen so as to simplify the values given in the table.
However, in order to verify the anomaly consistency conditions
such as $Tr\, Q = 8 Tr\, Q^3$, we must rescale the charges
appropriately. For instance, all values of $Q_1$ must be divided
by $\sqrt 2$.

\begin{table}
\begin{center}
\begin{tabular}{|c|c|c|c|}
\hline
$Sector $
& $SO(10)\times SU(5)$ & $Q_1$ &  $Q_2$   \\
\hline
$  U_1  $  &     (45,0) &   0  &   0  \\
\hline
& (1,5)   &    -1   &  1  \\
\hline
& $(1,\ov{5})$    &  1  &  -1 \\
\hline
\hline
$(\theta,A)$   & $2(16,1)$   &  -1/2  &  0\\
\hline
\hline
$(\theta,\ov{A}+\ov{S})$& $(10,1)$   &  -1/4  &  5/4  \\
\hline
& $(10,1)$ & 1/4 & -5/4 \\
\hline
& $(1,\ov{5})$ & -1/4 & 1/4 \\
\hline
& $(1,5)$ & 1/4 & -1/4 \\
\hline
& $(1,1)$ & 3/4 & 5/4 \\
\hline
& $(1,1)$ & -3/4 & -5/4 \\
\hline
\hline
$\omega$ & $ 3(1,\ov{10})$   & -1/2 & 1/2  \\
\hline
& $3(1,5)$   & -1/2 & -3/2  \\
\hline
& $3(1,1)$   & -1/2  & 5/2   \\
\hline
 & $(1,10)$   & 1/2   & -1/2   \\
\hline
& $(1,\ov{5})$   & 1/2   & 3/2  \\
\hline
 & $(1,1)$   & 1/2   & -5/2   \\
\hline
$\theta\omega$   & $3(16,1)$   &  -1/2  &  0 \\
\hline
 & $(\ov{16},1)$   &  1/2  &  0 \\
\hline
\end{tabular}
\end{center}
\caption{Particle content of the model of section 3.3.}
\label{t33}
\end{table}

\section{General aspects of orbifold GUTs}

There are many properties of the string GUTs presently described that
are general and apply to any  orbifold GUT. In fact some of them
are valid even for any level-two string GUT, independently of the
string-building method. Some of these general properties were spelled out in
ref.\,\cite{afiu}. In the present section we further elaborate on the
arguments to show that this class of string GUTs is very much
restricted both in particle content and couplings.
One of the main sources of constraints is
the fact that any massless particle must fulfill the condition
\begin{equation}
N_x\ +\ h_F\ +\ E_0\ =\ 1
\label{cons}
\end{equation}
$N_x$ is a bosonic oscillation number. $h_F$ is the contribution
to the left-handed conformal weight coming from the gauge coordinates.
$E_0$, given in (\ref{Evac}), is the contribution of the internal
twisted vacuum. There is a finite number of possibilities
for $E_0$, each one associated to a twist vector
$(v_1,v_2,v_3)$ (see Table 1 in ref.\,\cite{afiu}). On the other hand,
in our type of constructions, the gauge coordinates generically realize an
 affine Lie algebra together with a coset conformal field theory (CFT)
with conformal dimensions such that $c_F=c_{KM} + c_{coset}=16$.
Moreover, $h_F= h_{KM} + h_{coset}$, where $h_{KM}$ only
depends on the level of the algebra and on the gauge quantum numbers of
the particle according to
\begin{equation}
h_{KM}(R_1, R_2, \cdots )\ =\ \sum _a {{C(R_a)}\over {k_a+\rho_a}}
\label{formo}
\end{equation}
Here $C(R_a)$ is the quadratic Casimir of the representation $R_a$,
$k_a$ is the level and $\rho _a$ is de dual Coxeter number. The sum runs over
the different gauge groups.

\subsection{General properties of $SO(10)$ orbifold GUTs}

Using the constraint (\ref{cons}) and (\ref{formo}) it follows that
fields transforming like a $54$ can only be present in the
untwisted sector while adjoint $45$s can in principle exist
either in the untwisted sector or else in twisted sectors corresponding
to twist vectors $(0,\frac 14,-\frac 14)$ of $Z_4$ type
or $(0,\frac 16,-\frac 16)$ of $Z_6$ type \cite{afiu}.
The fact that a $54$ can only appear in the untwisted sector is easily
demonstrated since $h_{KM}(54)=1$ for $SO(10)$ at level two. Then
eq.(\ref{cons})
can only be verified for $E_0=0$ which means untwisted sector for symmetric
orbifolds. In the case of an adjoint $45$, $h_{KM}(45)= 4/5$ and hence
in principle any twist vector with $E_0\leq 1/5$ leaves room
enough for massless $45$s in the corresponding twisted sector. The unique
twists with $E_0\leq 1/5$ are precisely the $Z_4$ and $Z_6$ twists
mentioned above which have $E_0=3/16$ and $E_0=5/36$ respectively.
This is why in ref.\,\cite{afiu} we left open the possibility of finding
models with $45$s in these twisted sectors. We will now show that with a
little more effort this possibility can be ruled out to reach the
important conclusion that GUT-Higgs fields in orbifold $SO(10)$
string GUTs can only appear in untwisted sectors. This in turn has dramatic
consequences for the possible couplings of $45$s and $54$s. To prove
the above statement, it is useful to consider in more detail
the conformal field theory aspects of the gauge sector.
This will also allow us to understand other general properties of these
$SO(10)$ theories.

As we recalled in the introduction, $SO(10)$ string models obtained from
continuous Wilson lines are continuously connected to level-one
models with gauge group factors $SO(2N)\times SO(2M) \times G$,
with $N \geq M \geq 5$. The $45$s or $54$s can
only arise from level-one representations $(2N,2M)$ which have
conformal weight one for any $N,M$. Thus, neither these representations nor
the induced $45$s or $54$s can be present in any twisted sector.
The same is true, of course, for any $SO(10)$ string-GUT constructed
from flat directions in a level-one model. Thus,
$45$s in a twisted sector could only occur in a model which is not
continuously connected to level one.

Of the three methods considered in ref.\,\cite{afiu}, only that based on a
permutation action does not necessarily produce level-two models
continuously connected to level one. In the permutation method, $SO(10)$
at level two is obtained by starting from an orbifold with level one
gauge group $SO(2N)\times SO(2M)\times G$, $M\geq N\geq 5$. The permutation
$\Pi$ exchanges $L$ gauge coordinates inside $SO(2N)$ with another $L$
inside $SO(2M)$. As described in the previous section, massless states can
originate in permuted twisted sectors like $\theta_{\Pi}$ or
non-permuted like $\theta$. The latter cannot contain $45$s because their
particle content is a simple GSO-like projection of the level
one model that cannot contain $(2N,2M)$ multiplets in twisted sectors. Thus,
$45$s can only possibly exist in permuted sectors. Let us then concentrate
on this possibility.

{}From our previous results we know that eq.(\ref{cons}) is equivalent to
\begin{equation}
N_L\ +\ {1\over 2}(\ov{Q}+\ov{A})^2\ +\
{L\over {16}}\ +\ E_0\ =\ 1
\label{cuan}
\end{equation}
where $A$ is a generic gauge embedding in $\Lambda_{16}$.
Here $N_L=N_x+N_F$ contains the oscillator number $N_x$ of the bosonic
string coordinates and the oscillator number $N_F$ of the
gauge coordinates. The quantum numbers of the state are encoded in the
shifted $\Lambda_I^*$ momenta $(\ov{Q}+ \ov{A})$ and in $N_F$.
In particular, choosing to embed $SO(10)$ in the first five entries of
the shifted lattice, a $45$ must correspond to solutions of
eq.(\ref{cuan}) of the form
\begin{eqnarray}
& N_L=0\ \ ;\ \ (\ov{Q}+\ov{A})\ =\ (\underline{\pm \s2,
  \pm \s2,0,0,0}, \, Z;Y) & \\
& N_L=N_F=1/2\ \ ;\ \ (\ov{Q}+\ov{A})\ =\ (0,0,0,0,0,Z;Y) &
\label{soluc}
\end{eqnarray}
where $Z=(Z_6, \cdots, Z_{2L})$ and $Y=(Y_{2L+1}, \cdots, Y_{16})$.
The states with $N_F=1/2$ are the permuted oscillator states needed
to complete the $45$. The $Z$ and $Y$ vectors must coincide in both
types of states to ensure that the full $45$ has the same quantum
numbers under the rest of the gauge group.

The explicit solutions in (\ref{soluc}) are such that
\begin{equation}
N_L\ +\ {1\over 2}(\ov{Q}+\ov{A})^2\ \geq \ {1\over 2}
\label{eso}
\end{equation}
Combining this result with eq.(\ref{cuan}) implies
\begin{equation}
{L\over {16}}\ +\ E_0\ \leq \ {1\over 2}
\label{lim}
\end{equation}
As mentioned before, the only twisted sectors in which a $45$ has any
chance to appear have either $E_0=3/16$ or $E_0=5/36$. In either case,
eq.(\ref{lim}) can only be fulfilled for $L \leq 5$. Thus, only models
with permutations of five gauge coordinates can possibly work
and necessarily $Z=0$. In a $Z_4$ sector ($E_0=3/16$), the inequality in
(\ref{lim}) is saturated so that $Y=0$ whereas $Y^2=14/144$ in a
$Z_6$ sector. Since in this case $h_{KM}=\frac 45 + \frac{Y^2}2$, eq.
(\ref{cons}) in both sectors requires that the
$45$ be accompanied by a coset field with $h_{coset}=1/80$.

To proceed we need to further investigate the left-handed CFT
associated to the gauge degrees of freedom. Since we have
$L=5$ it is now clear that the relevant CFT has the structure
\begin{equation}
SO(10)_2\ \otimes \ \left( {{SO(10)_1\times SO(10)_1}\over {SO(10)_2}} \right)
\otimes \ G_Y   \ .
\label{coset}
\end{equation}
The central charges associated to each of the three factors is
$c_{SO(10)}=9$, $c_{coset}=1$ and $c_{Y}=rank{G_Y}=6$. In terms of these
theories we can write
\begin{equation}
h_F= h_{SO(10)}\ +\  h_{coset} \ +\ {{Y^2}\over 2}
\label{nuev}
\end{equation}
Thus, a particle will have contributions to its left-handed gauge
conformal weight coming from its $SO(10)$ and $G_Y$ quantum numbers
and from the $c=1$ coset.

We thus see that in building level-two $SO(10)$ string GUTs the coset theory
${\cal C}= SO(10)_1\times SO(10)_1 / SO(10)_2 $
plays an important r\^ole. Interestingly enough, this coset
has a remarkably simple CFT structure (which is not the case for the
equivalent coset in $SU(5)$ unification). Indeed, as explained
in ref.\,\cite{gins} , ${\cal C}$ is an element in the series
of $c=1$ models of type $SO(N)_1 \times SO(N)_1 /SO(N)_2$ that correspond
to $Z_2$ orbifolds of a free boson compactified on a torus
at radius $r=\sqrt{N} /2$. The conformal weights of winding and
momenta states of the circle compactification at this radius are given
by the simple expression
\begin{equation}
h_{m,n}\ =\ {1\over 2}
\left( {m\over {2(\sqrt{N} / 2)}}+n{{\sqrt{N}}\over 2}
\right)^2\ =\
{1\over {8N}}(2m+nN)^2 \ .
\label{gins}
\end{equation}
In our case $N=10$ and the weights in the circle compatification
are of the form $q^2/20$, $q\in {\bf Z}^+$. There are also $Z_2$ twist and
excited twist operators giving rise respectively to states of conformal
dimensions $\frac1{16}\,$ mod 1 and $\frac 9{16}\,$ mod 1.
The particles in our $SO(10)$ string GUT will have associated some of
the possible coset weights. No weight is as small as
$h_{coset}=1/80$, which is the required contribution for a massless
$45$ in the $Z_4$ or $Z_6$ sectors. Therefore, we conclude that
there cannot be $45$s in any twisted sector.

The fact that the GUT Higgs fields in $SO(10)$ string GUTs are necessarily in
the untwisted sector has important consequences for the structure of the
couplings in the theory. To begin, we can immediately show that there
will be just one GUT Higgs, either a $54$ or a $45$, associated
to one and only one of the three untwisted sectors. Indeed, in all the level
two $SO(10)$ constructions the starting level-one model has generic
gauge group $SO(2N)\times SO(2M)\times G$, with $M\geq N\geq 5$. To produce
such a group, the gauge embedding must have the general form
$A=(\oh, \oh, \cdots, \oh, \, 0,0, \cdots, 0, \, A_{M+N+1}, \cdots, A_{16})$
modulo elements of the $\Lambda_{16}$ lattice. Representations
transforming as $54$ or $45$ in the untwisted sector correspond
to specific subsets of lattice vectors of the form given in (\ref{pten}).
Since $P_{10,10} \cdot A=\oh$ modulo integer,
for these particles to survive the generalized GSO projection, they
must be associated to a complex plane which is twisted only by order-two
twists. This means that the internal twist vector
must be of the form $(v_1,v_2,\oh)$. This is the case of the
$Z_4$, $Z_6$, $Z_8$, $Z_{12}$, $Z_2\times Z_2$,
$Z_2\times Z_4$ and $Z_2\times Z_6$ orbifolds. All of these have a
complex plane which suffers only order-two twists and hence may have a $45$
or a $54$ in the corresponding untwisted sector. All other symmetric
orbifolds are ruled out for the purpose of $SO(10)$ model-building.
Finally, in the level-two $SO(10)$ model the $54$ and the $45$ cannot
both survive at the same time in the mentioned untwisted sector since
one of them is necessarily projected out due to the opposite
properties that these representations have under permutations of the
underlying $SO(10)\times SO(10)$ structure. In the case of a level-one
$SO(10)^2$ model continuously connected to the level-two
$SO(10)$ GUT, only a $54$ may be present in the massless sector since,
as discussed in \cite{afiu}, the $45$ is swallowed by the Higgs mechanism.

We thus see that the structure of symmetric orbifold string $SO(10)$ GUTs
is quite characteristic. The GUT-Higgs lives in one of the three
untwisted sectors, say in the third $U_3$. This very much constraints
its couplings. In particular, it is well known that the only
Yukawa couplings among purely untwisted particles is of the type
$U_1U_2U_3$. Also point-group invariance obviously forbids couplings of type
$UUT$. Thus, necessarily couplings of type
$54^3$ or $X54^2$, $X45^2$ are absent ($45^3$ couplings are anyway absent
due to its antisymmetric character). In fact, couplings of type
$45^n$, $54^n$ are forbidden for arbitrary $n$ due to the three R-symmetries
associated to each of the three complex planes in any orbifold.

There are also some conclusions that may be drawn {\it independently of the
string-construction method} and apply to methods like asymmetric
orbifolds or the fermionic construction. We already remarked that for
the $54$, $h_{SO(10)}=1$. Then, eqns. (\ref{cons})
and (\ref{nuev}) necessarily imply that a $54$ of $SO(10)$
cannot have any extra gauge quantum numbers,
otherwise it would be superheavy. Furthermore, $h_{coset}=0$, so that
its transformation properties under the coset CFT are
trivial. This implies the absence of couplings of type
$X(54)^2$ or (in theories with more than one $54$) $X(54)(54)'$, where
$X$ is any $SO(10)$ singlet field. Indeed, if $X$ is charged under some
other gauge symmetry, this gauge symmetry will forbid those couplings
since the $54$s are neutral. If on the other hand $X$ is neutral, it
necessarily has a non-trivial coset CFT structure.
In this second case the couplings also vanish because the corresponding
correlators should vanish given the trivial coset structure of the $54$s.
Notice however that a $54$ might have discrete (R-symmetry) quantum numbers
coming from the right-moving factor of its vertex operator.

\subsection{General aspects of $SU(5)$ orbifold GUTs}

In ref.\,\cite{afiu} it was found that the constraint (\ref{cons})
reduced the number of possibilities for the location of adjoint $24$s in the
different twisted sectors, since $h(24)=5/7$.
In particular, it was found that a $24$ could not
possibly be located in twisted sectors with
twist vectors: $(\frac 13,\frac 13, -\frac 23)$, $(\frac 14, \frac 14 ,-\oh)$,
$(\frac 16 ,\frac 13 , -\oh)$ and $(\frac 18 ,\frac 38, -\oh)$.
Arguing along the lines of the
previous section, we now show that actually $24$s cannot appear in
any twisted sector except in those of $Z_4$ and $Z_6$ type with
$(0,\frac 14, -\frac 14)$ and $(0, \frac 16, -\frac 16)$.
Furthermore, we will also show that $24$s cannot possibly appear
in the untwisted sectors of several Abelian orbifolds.

In the string GUT-building method in which the $SU(5)$ GUT is continuosly
connected through a flat direction to a level-one group such as
$SU(5)\times SU(5)$, the $24$s have their origin in representations
of type $(5,{\bar 5})$. Now, since such fields have $h_{SU(5)^2}=4/5$,
only twisted sectors with $E_0 \leq 1/5$ can contain a $24$.
Only the $Z_4$ and $Z_6$ twisted sectors mentioned above have this
property ($E_0=3/16$ and $E_0=5/36$ respectively).
It is now easy to prove that actually
only these twisted sectors may contain $24$s even if we
consider the permutation method which is not necessarily
continuously connected to level one.
Indeed, notice that we can still use formulae (\ref{cuan}) and (\ref{lim})
in this $SU(5)$ case, since the $SU(5)$ roots are a subset of those
in eq. (\ref{soluc}). Hence, we are again confined to the $Z_4$ and
$Z_6$ twisted sectors with $E_0\leq 3/16$.
Unlike the $SO(10)$ case, we cannot further argue about the absence
of $24$s in these two sectors since now $h(24)=5/7$ and
values of $h_{coset}$ larger than $(1-E_0-5/7)$ are allowed.
This leaves room for $24$s in these twisted sectors.
In fact, a $Z_2\times Z_4$ orbifold example of a $SU(5)$ model
with $24$s in a $Z_4$ sector was presented in ref.\,\cite{afiu}.

The alternative is having the $24$s in the untwisted sector. Let us then study
whether all orbifolds may have $24$s in the untwisted sector and, if so,
how many. The untwisted $24$s have their origin in chiral fields
of type $(5,\ov{5})$ or $(\ov{5},5)$ in the underlying $SU(5) \times SU(5)$
level-one structure (the group may be bigger but must contain
$SU(5)^2$). These representations have $P^2=2$ weights of the
form $P=(\underline {1,0,0,0,0}, \, \underline {-1,0,0,0,0}, \,
0,0,0,0,0,0)$ and $-P$ respectively.
For a model to have one $24$ in the untwisted
sector, its level-one ancestor must contain {\it at least two fields} of this
type in its own untwisted sector.
The reason is that one of these fields (or a combination of both)
dissappears in the process of going from the underlying level-one model
to the level-two $SU(5)$ GUT. Indeed, if the transition occurs by
giving vevs to one of these fields, one of them is Higgssed away. Thus,
two of these fields are needed so that one of them is left over. If level
two is reached by permutation modding, only half of the weights $\pm P$
(either symmetric or antisymmetric combinations) will survive the
projection, so that the final effect is similar.

We can see that there are just two ways to have two such fields. One
possibility is having an orbifold in which there is a degeneracy factor
$2$ or $3$ coming from the right-movers.
There are only three such orbifolds:$Z_3$ (multiplicity 3 for any
untwisted chiral field) and $Z_4,Z_6'$ (multiplicity 2 for the untwisted
fields in the first two complex directions). The second possibility
is having an orbifold with one of the complex planes feeling only order-two
twists (as in the $SO(10)$ case).
In this case, $P$ and $-P$ are allowed in the untwisted sector, since
a particle with weight $P$ as above has $(\pm P)\cdot A=1/2\,$ mod 1
($A$ is the shift in the underlying level-one theory).
Therefore, in order-two complex planes if there is
a $(5,\ov{5})$ there is also a $(\ov {5},5)$ and a
$24$ survives in the level-two $SU(5)$ model. Since for
orbifolds other than $Z_3$ ,$Z_4$ and $Z_6'$ a given weight in the gauge
lattice appears only once in the untwisted sector, we conclude that
the possibilities for obtaining at least one $24$ in the
untwisted sector are exhausted.

The above discussion allows us to rule out a number of orbifolds for the
purpose of constructing $SU(5)$ GUTs. Orbifolds without an order-two
plane or different from $Z_3$ and $Z_6'$ cannot lead to $24$s (unless they
arise from the order 4 or 6 twisted sectors that we mentioned above).
This eliminates orbifolds based on $Z_7$, $Z_8'$, $Z_3\times Z_3$ and
$Z_2\times Z_6'$. In fact, the $Z_3$ orbifold, can
also be discarded because if it has $(5,\ov {5})$s
in its untwisted sector, it automatically has vanishing net number of
fermion generations. We will spare the reader the proof of this statement.
It can be shown explicitly by considering the most general shifts and Wilson
lines compatible with the presence of these representations in the untwisted
sector. Thus, we conclude that for any $SU(5)$ symmetric orbifold GUT,
if the $24$s reside in the untwisted sector (by far the commonest case)
there is {\it only one} such $24$ associated to one of the three
untwisted sectors. The general situation concerning the possibilities offered
by symmetric $Z_N$ and $Z_M\times Z_N$
orbifolds for the construction of $SO(10)$ and $SU(5)$ GUTs
is summarized in Tables \ref{porron-a} and \ref{porron-b}.
Notice that only even-order orbifolds can give rise to string GUTs.

\begin{table}
\footnotesize
\begin{center}
\begin{tabular}{|c|c|c|c|c|c|c|c|c|c|}
\hline
$GUT $
& $Z_3$ & $Z_4$ &  $Z_6$  & $Z_6'$ & $Z_7$ & $Z_8$ & $Z_8'$ & $Z_{12}$ &
$Z_{12}'$ \\
\hline
$SO(10)$  & -- & $U$ & $U$ & -- & -- & $U$ & -- & $U$ & -- \\
\hline
$SU(5)$ & --   & $U$  &  $U$  & $U$ & -- & $U,T_4$ & -- & $U,T_6$ & $T_4$  \\
\hline
\end{tabular}
\end{center}
\caption{$Z_N$ orbifolds allowing for the construction of
string GUTs. The table indicates whether the GUT-Higgs are in
the untwisted sector (U) or in some order-four or six twisted
sector ($T_{4,6}$).}
\label{porron-a}
\end{table}

\begin{table}
\footnotesize
\begin{center}
\begin{tabular}{|c|c|c|c|c|c|c|c|c|}
\hline
$GUT $
& $Z_2\times Z_2$ & $Z_3\times Z_3$ &  $Z_2\times Z_4$  & $Z_2\times Z_6'$ &
$Z_2\times Z_6$  & $Z_4\times Z_4$ & $Z_3\times Z_6$ & $Z_6\times Z_6 $
\\
\hline
$SO(10)$  & $U$ & -- & $U$ & -- & $U$ & -- & -- & --  \\
\hline
$SU(5)$ & $U$   & --  &  $U,T_4$  & -- & $U,T_6$  & $T_4$ & $T_6$ & $T_6$  \\
\hline
\end{tabular}
\end{center}
\caption{$Z_M\times Z_N$ orbifolds allowing for the construction of
string GUTs. The table indicates whether the GUT-Higgs are in
the untwisted sector (U) or in some order-four or six twisted
sector ($T_{4,6}$).}
\label{porron-b}
\end{table}

As it happened in $SO(10)$, the fact that $24$s must belong either to the
untwisted or else to $Z_4$ or $Z_6$ twisted sectors has important implications
for the couplings of the GUT-Higgs. If the $24$ is in the
untwisted sector, $24^n$ couplings will be forbidden by the three
R-symmetries of the orbifold, and the same will be true for couplings of
$X(24)^2$ type, $X$ being any singlet particle. If, on the other hand, a
$24$ is present in a $Z_4$ ($Z_6$) sector, couplings of type
$24^{4n}$ ($24^{6n}$) are in principle allowed by
point group invariance. However, if the
$24$ is continuously connected to a $(5,\ov{5})$ representation
of an underlying $SU(5)\times SU(5)$ theory, then this underlying symmetry
will force the lowest dimension self-coupling to be of order 20 (30).
In any case cubic self-couplings $24^3$ as in the model of ref.\,\cite{dgs}
are forbidden. Notice however that the $24$s can couple to $5$ and $\ov{5}$s
in the theory, it is only self-couplings that are so much restricted.

\subsection{Summary of selection rules}

The above discussion may be summarized in a series of selection rules
which we collect in this subsection for the phenomenologically
oriented reader. There are some properties that are {\it general}
and apply to any level-two string GUT, whatever the construction
technique used. These are the following:

{\bf i.} All superpotential terms have $dim\geq 4$ (i.e. no mass terms).

{\bf ii.} At level two the only representations that can be present
in the massless spectrum
are: 5, 10, 15, 40 and 24 of $SU(5)$; 10, 16, 45 and 54-plets of $SO(10)$.

{\bf iii.} A $54$ of $SO(10)$ cannot be charged under any other
gauge group.

{\bf iv.} There cannot be couplings of type $X(54)^2$ or, if several
$54$s exist, $X(54)(54)'$, $X$ being any $SO(10)$-singlet chiral field.

The first two rules were already discussed in refs.\,\cite{fiq, afiu},
whereas the last two are discussed above.

If we further focus on the particular case of string GUTs
obtained from {\it symmetric orbifolds}, we can add the extra rules:

{\bf v.} $SO(10)$: There is only {\it either} one $54$ or one $45$.
They do not have any selfcouplings $54^n$, $45^n$ to any order.
Couplings bilinear in the GUT-Higgs such as
$(54)^2 XY \cdots $, $(45)^2XY \cdots$, are forbidden.
The $45$ or $54$ do on the other hand couple in general to Higgs $10$-plets.

{\bf vi.} $SU(5)$: There can be more than one $24$
only in a few orbifold
models containing particular $Z_4$ or $Z_6$ twists. Otherwise, there can be
only one $24$ with no self-couplings $24^n$ to any order.
Adjoints in twisted sectors can in principle have couplings
$(24)^{4n}$ or $(24)^{6n}$.

It is important to remark that properties {\bf v} and {\bf vi} apply only to
{\it symmetric orbifold  GUTs}. It cannot be concluded that all
string GUTs have this structure. Properties {\bf i} to {\bf iv} are general.

It is compelling to compare the resulting string GUT structure
with some $SO(10)$ models found in the recent SUSY-GUT literature.
In refs.\,\cite{bb1,bb2}, $SO(10)$ models in which doublet-triplet
splitting is effected by a version of the
Dimopoulos-Wilczek mechanism are considered.
They involve three adjoints, one 54 and a pair
$(16+\ov{16})$ with judiciously chosen couplings. This complicated
arrangement is required to guarantee doublet-triplet splitting while
avoiding Goldstone bosons and extra massless fields beyond those of the
MSSM. Refs.\,\cite{bb3, bm} study $SO(10)$ models
whose spectrum is claimed to be inspired by explicit
fermionic construction of string GUTs (in fact, the particular string
$SO(10)$ models referred to turned out to be inconsistent
due to lack of world-sheet supersymmetry).
What the authors actually do is to dispose of the $54$ while
trying to acomplish all the phenomenological goals with one single $45$,
and a pair $(16 + \ov{16})$. As they stand, all these models would have
difficulties to work in the string context since they violate the first of
the above selection rules by including explicit mass terms in the
superpotential.

In ref.\,\cite{hr} an attempt is made to get rid of explicit mass
terms and to use only dimension four potential terms.
The structure required to make
this work while maintaining doublet-triplet splitting, no Goldstone
bosons and a massless sector identical to that of the MSSM, is quite
contrived. Six adjoint $45$s, two $54$s, a pair $(16 + \ov{16})$ and
extra singlets, along with particular
superpotential couplings, are invoked in this scheme.
Such a multiplicity of GUT-Higgs fields is
rather unexpected in explicit string models (we just saw how in the case of
symmetric orbifolds only one GUT-Higgs is obtained). An extra problem may
concern the couplings. For example, a coupling of type $X(54)(54)'$ is used
($S_2S'S$ in the notation of \cite{hr}) which is strictly forbidden
in any possible level-two string GUT. This is a small detail which could
probably be easily cured. What seems difficult is to simultaneously procure
doublet-triplet splitting
and the MSSM massless content while using only $dim\geq 4$ couplings.
We think that this is a generic problem for $SO(10)$ GUTs and not solely
for those based on an underlying string theory.

\section{Four-generation and three-generation models}

In this section we present examples of string GUTs with reduced number of
generations (3 or 4). Models with an even number of net generations are
the commonest in our orbifold constructions.
Qualitatively speaking, what happens is that the number of
generations depends on the multiplicity of each twisted (and untwisted)
sector. The multiplicity of a twisted sector is in turn given by the number
of fixed points which is very often an even number for even orbifolds,
leading to even number of generations.
This is an oversimplification since there are even-order orbifolds with
some twisted sectors having odd multiplicities. Furthermore, the
degeneracy factors can be modulated to some extent through the
addition of Wilson lines. However, the above remark
reflects what actually occurs in explicit models.

Consider in particular the case of $SO(10)$ string GUTs.
As we have explained, these can only be realized
in orbifolds in which one of the three complex planes has at most
an order-two twist. This reduces the list of possibilities just to the
even orbifolds shown in Tables \ref{porron-a} and \ref{porron-b}. Since
we are dealing with strings based on the $Spin(32)/Z_2$ lattice, there are no
spinorial vectors with $P^2=2$ and hence $16$s of $SO(10)$ can never arise
in the untwisted sector, all generations must originate in twisted sectors.
Thus, the replication of generations results merely from degeneracy factors
in different twisted sectors that are even most of the time.
We have not carried out a full computer search of
three-generation $SO(10)$ string GUTs but just
an exploratory scan in the lowest order orbifolds ($Z_4$, $Z_6$,
$Z_2\times Z_2$ and $Z_2\times Z_4$) and we have not found three-generation
models. However, we do not think that
this is a general feature of string theory
and probably three-generation examples can be found. After all, the first
Calabi-Yau compactifications historically found all had even number of
generations, it took some time to produce three-generation examples!.
Notice in this context that the three-generation $SO(10)$ models reported in
\cite{lyk,cleav}
were later on withdrawn due to lack of world-sheet supersymmetry.

The situation of $SU(5)$ GUTs is slightly different. According to Tables
\ref{porron-a} and \ref{porron-b}, the list of possible orbifolds is somewhat
larger. Furthermore, unlike the $SO(10)$ case, there can be
generations in the untwisted sector since $(10 + \ov{5})$ representations
need not come from $\Lambda_{16}$ spinorial weights.
Thus, in principle the $SU(5)$ GUT construction is more flexible with
respect to the generation number. For instance, if there were four
generations from the twisted sectors, there could be an antigeneration
in the untwisted sector to adjust three net families. An example of
this sort will be shown below. However, it turns out that these
attempts to look for three generations lead to models considerably
more complicated than the four-generation models that we obtain.
Therefore, we think that it is interesting to first examine some aspects of
four-generation models. In particular, we want to discuss
the Yukawa coupling structure of the four-generation
$SO(10)$ GUT called ``example 1'' in ref.\,\cite{afiu},
due to its relative simplicity and also because there are many similar
$SO(10)$ string GUTs that can be obtained with the present techniques.
We also briefly discuss a few aspects of the
four-generation examples in section 3.
A recent study of the phenomenological viability of four-generation SUSY
models is carried out in ref.\,\cite{masiero}.

\subsection{A four-generation $SO(10)$ GUT}

We consider the first example discussed in chapter 4 in ref.\,\cite{afiu}.
This model can in fact be built by the three orbifold methods:
continuous Wilson lines, permutation modding or flat directions.
In ref.\,\cite {afiu} it is constructed through continuous Wilson lines
on a $Z_2\times Z_2$ orbifold with cubic $SO(4)^3$ torus lattice.
The $Z_2\times Z_2$ twists are such that the twisted sectors have
reduced multiplicities (see \cite{afiu} for details).
The gauge group is $SO(10)_2\times SO(8) \times U(1)^2$ where the GUT is
realized at level two and the rest at level one. The complete
chiral field massless spectrum is shown in Table \ref{t51}.

\begin{table}
\begin{center}
\begin{tabular}{|c|c|c|c|c|}
\hline
$Sector $
& $SO(10)\times SO(8)$ & $Q$ &  $Q_A$  & Notation \\
\hline
$  U_1  $  &     (1,8) &   1/2   &   1/2 &  \\
\hline
& (1,8)   &    -1/2   &  -1/2 &  \\
\hline
$  U_2  $ & (1,8)   & -1/2 & 1/2 &  \\
\hline
& (1,8)    &   1/2  &  -1/2  & \\
\hline
$   U_3  $ & (54,1)   & 0   &  0 & $\Phi _{54}$ \\
\hline
& (1,1)    &    0  & 0  & $X_0$ \\
\hline
& (1,1)    &  0  & 1  & $X_{01}$  \\
\hline
& (1,1)    &   1  & 0  & $X_{10}$\\
\hline
& (1,1)    &  -1  & 0  & $X_{-10}$ \\
\hline
& (1,1)    &  0  & -1  & $X_{0-1}$ \\
\hline
$\theta$   & $3(16,1)$   &  1/4  &  1/4  & $16_{\theta }^i$ \\
\hline
& $(\ov{16},1)$   &  -1/4  &  -1/4 & $\ov{16}_{\theta }$  \\
\hline
$\omega$   & $3(16,1)$   &  -1/4  &  1/4 & $16_{\omega }^i$ \\
\hline
 & $(\ov{16},1)$   &  1/4  &  -1/4 & $\ov{16}_{\omega }$ \\
\hline
$\theta\omega$ &    $ 4(10,1)$   & 0   & 1/2 &  $10_+^a$ \\
\hline
 & $4(10,1)$   & 0   & -1/2  & $10_-^a$ \\
\hline
      & $3(1,8)$   & 0   & 1/2  & \\
\hline
 & $(1,8)$   & 0   & -1/2 &  \\
\hline
   & $8(1,1)$   & 1/2   & 0 & $X_+^r$ \\
\hline
 & $8(1,1)$   & -1/2   & 0  & $X_-^r$ \\
\hline
\end{tabular}
\end{center}
\caption{Particle content and charges of the 4-generation $SO(10)$ model
of section 5.1.}
\label{t51}
\end{table}

The $U(1)$ with charge $Q_A$ is anomalous but this anomaly
is cancelled by the Green-Schwarz mechanism. The fact that $TrQ_A\not=0$
induces a one-loop effective Fayet-Illiopoulos term. Upon minimization of the
scalar potential this term will force some chiral field to acquire a vev
in order to cancel the D-term. Since $TrQ_A$ is possitive,
giving a vev to $X_{0-1}$ will be sufficient for this cancellation.

It is useful to recall some properties of the Yukawa couplings.
Generically denoting fields by their corresponding sector,
the allowed Yukawa couplings in the $Z_2\times Z_2$ orbifold
are of the form
\begin{equation}
U_1U_2U_3\ , \ U_2\theta \theta \ , \
U_1 \omega \omega \ , \ U_3 (\theta \omega)(\theta \omega
)\ , \ \theta \omega (\theta \omega )\
\label{permit}
\end{equation}
These are the only couplings allowed by the point-group and $H$-momentum
discrete symmetries. Among the allowed cubic
couplings a small subset could be actually forbidden by extra discrete
symmetries coming from space-group selection rules that
depend on details of the structure of fixed points and
space-group conjugacy classes. Since we just want
to point out a few qualitative features we will not take into account these
further restrictions. It is interesting to remark
that, in spite of the many possible couplings forbidden by
point-group and $H$-momentum selection rules, the distribution of the
particles in the different untwisted and twisted sectors is such that
many of the phenomenologically required couplings are indeed present.
The list of interesting Yukawa couplings in the model includes the following:

i) The singlets $X_0, X_{01},X_{10}$,$X_{-10},X_{0-1}$ in the $U_3$ untwisted
sector have couplings with the $SO(8)$ octets in the sectors $U_1$, $U_2$ and
$\theta \omega$. If some or all of these singlets get vevs (we already saw that
the Fayet-Illiopoulos term induces a vev at least for $X_{0-1}$), most or all
of this extra matter will dissapear from the massless spectrum. At the same
time
the $U(1)^2$ symmetry is spontaneously broken.

ii) The singlets $X_+^r$, $X_-^r$ in the $\theta \omega $ sector have couplings
to some $(16+ \ov{16})$ pairs:
\begin{equation}
h_{ri}X_+^r(16_{\omega }^i)(\ov{16}_{\theta })\quad  ; \quad
h_{rj}'X_-^r(16_{\theta }^i)(\ov{16}_{\omega })\
\label{sing}
\end{equation}
where $h,h'$ are Yukawa coupling constants of order unity.
These couplings may have several effects.
Some of these singlets $X_{+,-}$ could get a vev and give large masses to
one or two $(16+\ov{16})$ pairs. A second use of these couplings
is perhaps more interesting. We know that the breaking of the
$SO(10)$ symmetry down to the standard model requires vevs both
for the $54$ and a $(16+ \ov{16})$ pair. If one of the two $\ov{16}$,
say  $\ov{16}_{\theta }$, gets a vev, the right-handed neutrinos of
the three $16_{\omega }$ generations will combine with three of the eight
$X_+^r$ singlets and will dissapear from the massless spectrum.
Thus, this would solve the
neutrino mass problem for these three generations that would be naively
identified with the observed three generations. On the other hand, the fourth
net generation would be one of the three
$16_{\theta }^i$ fields (or a lineal combination).
Since they do not couple to the $\ov{16}_{\theta }$
field, their right and left-handed neutrinos can only have
Dirac type masses that could be large if their standard Yukawa coupling is
of order one. In principle this
could explain why LEP sees only three and not four
massless neutrinos, although of course a more detailed analysis would be needed
to reach a definitive conclusion. Such detailed analysis goes beyond the scope
of this paper in which a systematic study of the phenomenology of
the models is not pursued.

iii) Yukawa couplings of the standard type that can give usual Dirac masses to
the fermions are also present. The symmetries allow for the couplings:
\begin{equation}
h_{ija}16_{\theta }^i16_{\omega }^j10_-^a \quad  ; \quad
h_{ija}'\ov{16}_{\theta }^i \ov{16}_{\omega }^j10_+^a
\label{yuki}
\end{equation}

iv) The electroweak Higgsses $10_{+,-}$ couple to the fields $\Phi _{54}$,
$X_0$, $X_{01}$ and $X_{0-1}$ of the $U_3$ sector :
\begin{equation}
(10_+,10_-)\left(
\begin{array}{cc}
    X_{0-1} & X_0+{\Phi _{54}} \\
  X_0+{\Phi _{54}}   &     X_{01}
\end{array}
\right)
\left(
\begin{array}{c}
10_+ \\
10_-
\end{array}
\right)
\label{sliding}
 \end{equation}
These couplings are similar to those of the singlets in
$U_3$ to the $SO(8)$ octets mentioned above. There is however an
important difference in this case signaled by the presence
of couplings to the GUT-Higgs $\Phi _{54}$. If there is
a vev $\langle \Phi _{54} \rangle =v\,$diag$\,(2,2,2,2,2,2,-3,-3,-3,-3)$,
the $SO(10)$ symmetry breaks down to $SU(4)\times SU(2)\times SU(2)$. Then,
there are regions in the ``moduli-space'' of the fields $\Phi _{54}$,
$X_0$, $X_{01}$ and $X_{0-1}$ in which some electroweak multiplets
$(1,2,2)$ can be light and their $SO(10)$ partners, the  $(6,1,1)$s
are heavy. Thus, the possibility of doublet-triplet splitting exists but,
as already remarked in \cite{afiu} , it remains to understand
why the regions of moduli-space giving some massless doublets should be
dynamically preferred.
The existence of all the above phenomenologically required couplings is
non-trivial. However, a much more detailed analysis of the different couplings
and a better understanding of the doublet-triplet splitting would be needed.

Before closing this subsection let us briefly comment on the
phenomenologically interesting couplings of the four-generation
models constructed in section 3. The model in section 3.1
is an $SO(10)$ GUT with a $45$ and no $(16+\ov{16})$
pairs (see Table 1). The four $\ov{16}$ generations
in the $(\theta, V)$ sector have standard Yukawa couplings with
the first set of $10$-plets in the $({\theta }^2, 2V)$ sector.
The adjoint $45$ in the $U_3$ sector can potentially split
doublets from triplets due to its coupling to the two sets of Higgs
$10$-plets in the $({\theta }^2,2V)$ sector. In spite of the existence
of these interesting couplings, the absence of $(16+\ov {16})$
pairs makes impossible the breaking of the $B-L$ symmetry
down to the standard model and also renders the
right-handed neutrinos insufficiently massive.
The $SO(10)$ model in section 3.3
(see Table 3) has a $(16+\ov {16})$
pair and the breaking of the symmetry down to the standard
model can easily proceed. However, the standard fermion Yukawa
couplings and the coupling of the $45$ to the Higgs $10$-plets
are not present at the renormalizable level. Although the required
effective Yukawa couplings might appear from nonrenormalizable terms,
it is fair to say that the $SO(10)$ model with a $54$ discussed
above has a simpler structure.

The $SU(5)$ model in section 3.2 (see Table 2) has an interesting
Yukawa coupling structure. The $SU(5)$ $10$-plets belong to
sector $(\theta , V)$ and couplings to the
last of the $\ov{5}$-plets in the
$\theta ^2$ sector exist. These give masses to the $U$-type quarks.
Similar couplings exist for the $D$-type quarks.
Couplings of the $24+1$ representation in the $U_3$ sector to the
$(5+\ov{5})$s fields in the $\theta ^2$ sector do also
exist. These couplings potentially allow for doublet-triplet splitting.
Of course, these comments about these four-generation models
just reflect the most obvious aspects of their Yukawa coupling
structure. A more detailed analysis including the effects of
non-renormalizable couplings would be needed to gauge their
phenomenological viability.

\subsection{A $SU(5)$ string GUT with ``almost'' three generations}

Perhaps the most
efficient method for the purpose of model searching
is to start with a level-one model with replicated GUT groups
and then find flat directions in which the symmetry is broken down
to a diagonal level-two GUT. Here we present an $SU(5)$ GUT with adjoint
Higgsses obtained by taking flat directions  breaking a level-one
gauge group $SU(6)\times SU(5)$ to a level-two $SU(5)$.
This is a $Z_6$ orbifold generated by the twist vector
$(\frac 16, \frac 26, -\frac 36)$ acting on an $SU(3)^3$ compactification
lattice. The embedding in the gauge degrees of freedom is realized by a
shift:
\begin{equation}
V\ =\ {1\over 6}(1,1,1,1,1,1,-2,-2,-2,-2,-2,3,3,3,3,0)
\label{shif}
\end{equation}
This shift fulfills the usual level matching condition
$6(V^2-v^2)=even$.
The $SO(32)$ gauge symmetry is broken down to
$SU(6)\times SU(5)\times SO(8)\times U(1)^3$.
No Wilson lines are needed in this particular model. The massless chiral
spectrum shown in Table \ref{t52} is found following the usual
projection techniques.

\begin{table}
\begin{center}
\begin{tabular}{|c|c|c|c|c|}
\hline
$Sector $
   & $SU(6)\times SU(5)\times SO(8)$ & $Q_1$   &  $Q_2$
&   $Q_3 $       \\
\hline
$  U_1  $  &   $(6,1,1)$ &   1   &    0  &   1     \\
\hline
&   $(6,1,1)$ &   1   &    0  &   -1    \\
\hline
      &   $(\ov{6},\ov{5},1)$   &    -1   &  -1   &   0  \\
\hline
      &   $(1,5, 8_V)$ & 0   &    1  &   0   \\
\hline
$  U_2  $  &   $(1,\ov{5},1)$ &   0   &    -1  &  1  \\
\hline
&   $(1,\ov{5},1)$ &   0   &    -1  &  -1   \\
\hline
    &   $(\ov{6},1,8_V)$   & -1   &  0   &  0    \\
\hline
    &   $(1,10, 1)$ &   0   &    2  &   0    \\
\hline
    &   $(15,1, 1)$   & 2   &  0   &  0    \\
\hline
$  U_3  $  &   $(1,1,8_V)$ &   0   &    0  &  1   \\
\hline
&   $(1,1,8_V)$ &   0   &    0  &  -1   \\
\hline
    &  $(6,\ov{5},1)$   &   1   &  -1   &  0    \\
\hline
      &   $(\ov{6},5,1)$ &   -1   &    1  &    0  \\
\hline
 $\theta $    &
$12(6,1,1)$   &  -1  & 5/6    &  -1/2  \\
\hline
    &   $12(1,1,1)$  &  -2  & 5/6 &  1/2  \\
\hline
$\theta ^2$    &
$3(1,1, 8_{\bar s})$   &  -1  &  -5/6   & 1/2     \\
\hline
    &   $3(1,1,8_{\bar s})$  &  1  &  5/6 & 1/2     \\
\hline
    &   $6(1,1,8_s)$  &  -1  &   -5/6     &  -1/2  \\
\hline
    &   $6(1,1,8_s)$  &  1  &   5/6     &  -1/2  \\
\hline
    &   $3(6, 1,1)$   &  -1  &  -5/3   &    0  \\
\hline
    &   $6(\ov{6},1,1)$  &  1   &  5/3   &   0  \\
\hline
    &   $3(1,{\bar 5},1)$  &  2  &  2/3      &  0  \\
\hline
    &   $6(1, 5,1)$  &  -2  &  -2/3      &  0  \\
\hline
$\theta ^3 $    &
$4(1,1,1)$   &  0    &  5/2   & -1/2     \\
\hline
    &   $8(1,\ov{10},1)$  &  0  & 1/2   & -1/2
\\
\hline
    &   $4(1, 5,1)$  &  0  & -3/2  &  -1/2    \\
\hline
&   $8(1,1,1)$  &  0  & -5/2   &  1/2      \\
\hline
    & $4(1,10, 1)$   &  0    &  -1/2   & 1/2      \\
\hline
    &   $4(1,\ov{5},1)$  &  0  & 3/2   & 1/2    \\
\hline
\end{tabular}
\end{center}
\caption{Particle content and charges of the $SU(6)\times SU(5)$
model of section 5.2.}
\label{t52}
\end{table}

In the third untwisted sector, associated to the complex plane
with an order-two twist, there appear the multiplets
$(6,\ov{5},1)$ + $(\ov{6},5,1)$. There is an F-flat and D-flat
field direction corresponding to the vevs:
\begin{equation}
\langle (6^i,\ov{5}^a,1) \rangle \ =\ \langle (\ov{ 6}^i,5^a,1)\rangle  \ =
\ \nu \delta^{ia} \quad ; \quad i,a=1,\cdots , 5
\label{flat}
\end{equation}
with $i\not= 6$. These vevs break the $SU(6)\times SU(5)$ symmetry down to
$SU(5)_2\times U(1)$, where the $SU(5)_2$ is realized at level two.
The model so obtained constitutes an $SU(5)$ GUT. There is an adjoint $24$
in the $U_3$ untwisted sector arising from the chiral fields in
$(6,\ov{5},1)$ + $(\ov{6},5,1)$ that are not swallowed by the
Higgs mechanism.

The net number of $SU(5)$ generations is found by looking
for the $10$-plets in the spectrum. In the $\theta ^3$ twisted sector
there are four net $\ov{10}$ multiplets. However, there is an
extra massless $10$-plet from the untwisted sector so that there are
indeed three chiral $(\ov{10}+5)$ generations plus a number of
vector-like $(10+\ov{10})$ and $(5+\ov{5})$ multiplets. The
origin of this additional $10$-plet in the untwisted sector is as follows.
Naively we would say that there are in fact two $10$-plets in the untwisted
sector coming from the $(1,10,1)$ and $(15,1,1)$ fields (recall that
the latter decomposses under $SU(5)$ as $15=10+5$).
However, the following couplings between untwisted particles exist:
\begin{equation}
(6,\ov{5})(1,10)(\ov{6},\ov{5})\ \ \  ;\ \ \
(\ov{6}, 5)(15,1)(\ov{6},\ov{5})\
\label{mir}
\end{equation}
Once the $(6,\ov{5})$, $(\ov{6}, 5)$ acquire the vevs in eq. (\ref{flat}),
the antisymmetric piece of the $(\ov{6},\ov{5})$ combines with
a linear combination of the two $10$-plets inside $(15,1)$ and $(1,10)$
and becomes massive. Thus, only one $10$-plet from the untwisted sector
remains in the massless
spectrum and the net number of standard $SU(5)$ generations is indeed three.

Unfortunately, apart from this three generations and the vector-like matter
there is an extra ``exotic family'' in the massless spectrum.
As it is well known, the simplest
chiral anomaly-free combination in $SU(5)$ is $(10+\ov{5})$. The next to
simplest chiral anomaly-free combination is $(15+9\cdot \ov{5})$, where
this $15$-plet is the two index symmetric tensor (do not mistake it with the
other $15$-plet above which is the antisymmetric $SU(6)$ tensor). One of these
``exotic families'' is present in the massless spectrum of this model.
A $\ov{15}$ is indeed obtained from the symmetric components of the
$(\ov{6},\ov{5})$ fields and remains in the massless spectrum.
There is also a surplus
of nine $5$-plets that cancel the $SU(5)$ anomalies.
It would be interesting to study in detail to what extent this extra ``exotic
family'' could be made phenomenologically viable.
Under the standard model
group, $15=(6,1,-2/3)+(3,2,1/6)+(1,3,1)$ and hence we would have at least
an extra colour-sextet quark. Objects of this type have been considered
in the past \cite{quigsses} .
Apart from this potential problem, the structure of Yukawa
couplings is much less appealing than that found for the previous
four-generation $SO(10)$ model
(see ref.\,\cite{japs} for the general structure of $Z_6'$ Yukawa couplings).
For example, the $\ov{10}$-plets corresponding to
the three physical generations are in the $\theta ^3$ twisted sector. However
there are no Yukawa couplings of the type $\ov{10}\, 5 \, \ov{10}$, so
that, for instance,
the top-quark would have no Yukawas at this level. It is also not obvious
how doublet-triplet splitting could occur.


\section{Conclusions and Outlook}

In this paper we have further explored the
construction and patterns of 4-d
symmetric orbifold strings  whose massless spectrum
constitue $SO(10)$ and $SU(5)$ GUTs. From this study there arises
a very characteristic structure for this class of
string GUTs. We find that only certain even-order
Abelian orbifolds can be used to construct such models.
In particular, we have shown on general grounds that
 $SO(10)$ or $SU(5)$ GUTs cannot be obtained from
the $Z_3$, $Z_7$, $Z_3\times Z_3$, $Z_8'$ and $Z_2\times Z_6'$
orbifolds. String $SO(10)$ models can only be obtained
in the subset of even-order orbifolds that have
a complex plane rotated only by order-two twists.
They only contain either a single $54$ or a single $45$
that belongs in the untwisted sector corresponding to that
order-two twist. Due to their location  in an untwisted
sector, these GUT-Higgs fields behave very much like
string moduli.

In the case of string $SU(5)$ models the situation is slightly less
tight. The models usually contain a single adjoint $24$
located in an untwisted sector, which again leads to modulus-like
behaviour for this field. Multiple $24$s can however arise in a
restricted class of Abelian orbifolds that contain certain
twisted sectors of order four or six.
In both $SO(10)$ and $SU(5)$ string symmetric orbifolds the
couplings of the GUT-Higgs fields are very constrained.
In section 4 we list a number of selection rules for their couplings.
Some of these rules have more general validity
and apply to any possible level-two string GUT, and not only
to symmetric orbifolds.

We have also extended our previous analysis concerning
the permutation method to construct level-two models.
The models obtained through this method, unlike the other
two used in \cite{afiu} are not necessarily continuously connected
to level-one models. This offers new model-building possibilities.
In particular this method allows us to obtain $SO(10)$ GUTs
with adjoint $45$ Higgs fields (in the other two methods
only $54$s can be obtained). Our study of the permutation
method includes a careful study of the partition function
in order to obtain the appropriate modular invariance conditions
as well as the generalized GSO projectors.

In the context of symmetric orbifold GUTs it is natural
to obtain models with an even number of generations.
As we said above, only even-order orbifolds can be used
to construct GUTs and this typically leads to even degeneracies
in the twisted sectors which in turn implies an even number of
generations. To illustrate our method we have
presented several four-generation $SO(10)$ and $SU(5)$ models.
It is intriguing that many of the phenomenologically
required couplings, e.g. couplings giving
masses to right-handed neutrinos, couplings which could induce
doublet-triplet splitting, usual Yukawa couplings, etc., are indeed present.
Recently there has been some interest in the study of
four-generation versions of the SSM
\cite{masiero} . It could perhaps be interesting
to study in detail the possible phenomenological viability of
some of these four-generation string GUTs.
The closest we get to a three-generation model is
an $SU(5)$ example with an extra exotic
$(15+9\cdot \ov{5})$ ``family'', but its phenomenological properties
do not seem to be better than those of the four-generation models.
Although it indeed seems that even number of generations is the most
common situation in symmetric orbifold string GUTs, our search has
not been systematic and we cannot rule out at the moment the existence
of three-generation examples in this class of string GUTs.
A computer search for models could perhaps be appropriate.

The methods presented here can be generalized
in several ways. An obvious possibility is the
construction of level-two models based on {\it asymmetric } orbifolds.
Indeed, as remarked in \cite{afiu}, the strong constraints
coming from eq. (\ref{cons}) are much weaker in the case of
asymmetric orbifolds since, for instance, there can be cases in which the
right-movers are twisted but the left-movers are not. In this case
$E_0=0$ in eq. (\ref{cons}) and it becomes easier to find
GUT-Higgs fields in twisted sectors, leading to models
with multiple GUT-Higgses. Indeed, string GUTs with multiple
GUT-Higgs fields based on asymmetric orbifolds can be constructed
\cite{afiuup, erler}. The trouble is that very
little model-building, if any, has been done using
asymmetric orbifolds and it would be necessary to first develop
some theoretical tools. In particular, a good control
of the generalized GSO projectors for asymmetric
orbifolds would be necessary to find the complete massless
spectrum of each model. Although in principle this can
be extracted from the original literature \cite{nsv}
in practice this is a non-trivial task. This task
becomes even more involved in the presence of the extra
modding required to get the level-two model.
Furthermore, the addition of discrete Wilson lines in
asymmetric orbifolds turns out to be rather restricted
compared to the symmetric case \cite{imnq} , and this
makes the construction less flexible from the model-building
point of view.

A different, perhaps more promising, approach is the
derivation of level-two GUTs based on the Gepner or
Kazama-Suzuki constructions. The techniques employed for
symmetric orbifolds can be easily generalized to
start with a coset compactification of
the $Spin(32)/Z_2$ heterotic string with an $SO(26)$ gauge group
(plus possible additional $U(1)$s). The internal $c=9$ system in this
case is a tensor product of $N=2$ superconformal coset models.
As discussed in detail in refs.\,\cite{mond, fiqs},
different gauge groups and particle contents can be obtained by twisting
the $N=2$ superconformal models by discrete symmetries
while embedding these symmetries in the $SO(26)$
degrees of freedom through a shift. In this way, level-one
models with replicated groups are found as in the orbifold case.
Then, if the given coset compactification has an order-two
symmetry, it could be embedded through an order-two permutation of two
GUT group factors, taking care that left-right level matching
is mantained. The class of string GUTs obtained in this
way would be rather large and so will be the chances to find three
generation models.

The outstanding challenge in GUT model-building, with or without
strings, is the origin of doublet-triplet splitting.
The structure of the standard model is quite bizarre. There is
a chiral piece, the quark-lepton generations, and
a vector-like piece, the Higgs pair $H_1$, $H_2$. Even with
supersymmetry, the presence of the light Higgs fields is a mistery.
In the case of GUTs this mistery comes in the form of the doublet-triplet
splitting problem. But in the context of other level-one,
non-GUT, string models
the mistery still exists of why a set of Higgs fields remains light
anyhow (even though no splitting is needed). Thus, we do not think
that the situation of string GUTs concerning this issue
is conceptually very different from the situation of
other string level-one models. On the contrary,
as discussed in ref.\,\cite{afiu}, within the context
of string GUTs, doublet-triplet splitting
appears as a dynamical problem in which the crucial issue is the
quantum moduli space of the GUT-Higgs fields in the model.
Perhaps recent developements in the understanding of
non-perturbative phenomena in supersymmetric theories could
help in solving this problem.

In closing, let us comment on
level-two versus level-one 4-d strings.
One may have the wrong impression that
somehow level-one heterotic theories are more generic than
level-two or higher. In fact, one may argue for the contrary.
We remarked at the beginning of section two
(see also ref.\cite{fiq})
that the higher level theories are often continuously connected
to level-one theories. For {\it generic} values of some charged moduli
fields the gauge group has higher level and it is only
at particular points of moduli space that the gauge group is enhanced
to a bigger one realized at level-one. Thus, in this sense,
reduced rank models realized at higher levels
are more generic than level-one models.
Independently of its use to construct GUT theories,
level-two (and higher) string theories
could be relevant in realizing the low-energy physics.
In particular, it is conceivable that all or some of the
non-Abelian symmetries of the standard model could correspond
to higher level affine Lie algebras. We hope that the techniques
developed in \cite{afiu} and in the present article
will be useful to the general study of higher level
heterotic strings, independently of their use in
GUT model-building.

\bigskip
\bigskip

\centerline{\bf Acknowledgements}
\medskip
G.A. thanks the ICTP and the Departamento de F\'{\i}sica Te\'orica
at UAM for hospitality, and the Ministry of Education
and Science of Spain
as well as CONICET (Argentina) for financial support.
A.F. thanks the Departamento de F\'{\i}sica Te\'orica at UAM
for hospitality and support at intermediate stages of this
work, and CONICIT (Venezuela) for a research grant S1-2700.
L.E.I. thanks the CERN Theory Division for hospitality.
A.M.U. thanks the Government
of the Basque Country for financial support. This work has also been
financed by the CICYT (Spain) under grant AEN930673.

\end{document}